\documentclass[aps,amsfonts,prl,twocolumn,showpacs]{revtex4-1}

\usepackage{subfigure}
\usepackage{textcomp}
\usepackage{graphicx}
\usepackage{amssymb}
\usepackage{amsmath}
\usepackage{bm}
\usepackage{amsmath, amsthm, amssymb}
\usepackage{dsfont}
\usepackage[colorlinks=true,linkcolor=blue,urlcolor=blue,citecolor=blue]{hyperref}
\usepackage{multirow}
\usepackage{url}

\usepackage{pdfpages} 
\makeatletter
\AtBeginDocument{\let\LS@rot\@undefined}
\makeatother

\def\beq{\begin{equation}}
\def\eeq{\end{equation}}
\def\bea{\begin{eqnarray}}
\def\eea{\end{eqnarray}}

\begin{document}

\title{Anomalous low-frequency conductivity in easy-plane XXZ spin chains}

\author{Utkarsh Agrawal$^1$, Sarang Gopalakrishnan$^2$, Romain Vasseur$^1$, and Brayden Ware$^{1,3}$}
\affiliation{$^1$ Department of Physics, University of Massachusetts, Amherst, MA 01003, USA}
\affiliation{$^2$ Department of Physics and Astronomy, CUNY College of Staten Island, Staten Island, NY 10314;  Physics Program and Initiative for the Theoretical Sciences, The Graduate Center, CUNY, New York, NY 10016, USA}
\affiliation{$^3$ Rudolf Peierls Center for Theoretical Physics, Clarendon Laboratory, University of Oxford, Oxford OX1 3PU, UK}

\begin{abstract}

In the easy-plane regime of XXZ spin chains, spin transport is ballistic, with a Drude weight that has a discontinuous fractal dependence on the value of the anisotropy $\Delta = \cos \pi \lambda$ at nonzero temperatures. We show that this structure necessarily implies the divergence of the low-frequency conductivity for generic irrational values of $\lambda$. Within the framework of generalized hydrodynamics, we show that in the high-temperature limit the low-frequency conductivity at a generic anisotropy scales as $\sigma(\omega) \sim 1/\sqrt{\omega}$; anomalous response occurs because quasiparticles undergo L\'evy flights. For rational values of $\lambda$, the divergence is cut off at low frequencies and the corrections to ballistic spin transport are diffusive. We also use our approach to recover that at the isotropic point $\Delta=1$,  spin transport is superdiffusive with $\sigma(\omega) \sim \omega^{-1/3}$.
We support our results with extensive numerical studies using matrix-product operator methods.

\end{abstract}

\maketitle

Integrable models play a special part in our understanding of quantum many-body physics: on the one hand, they are among the few strongly interacting quantum systems for which exact results exist; on the other, their dynamics are special by virtue of their integrability. For example, integrable systems have extensively many conserved quantities~\cite{PhysRevLett.106.217206, PhysRevLett.110.257203, PhysRevLett.113.117202, PhysRevLett.115.157201} and stable ballistically propagating quasiparticles, unlike quantum chaotic systems. Although integrability is technically a fine-tuned property, many experimentally relevant one-dimensional models---such as the Hubbard, Heisenberg, and Lieb-Liniger models---are either exactly or approximately integrable~\cite{takahashi}. The dynamics of integrable and nearly integrable models have lately been extensively studied, both theoretically~\cite{Calabrese:2006, PhysRevLett.106.217206, PhysRevLett.110.257203, PhysRevLett.113.117202,PhysRevLett.113.117203, PhysRevLett.115.157201, 2016arXiv160300440I, 1742-5468-2016-6-064001,1742-5468-2016-6-064002,1742-5468-2016-6-064010,1742-5468-2016-6-064007,PhysRevB.89.125101, alba2017entanglement} and experimentally~\cite{kinoshita, gring, tang2018, erne2018, zundel2018}.

Although the exact dynamics of large integrable systems remains challenging, the recently developed framework of generalized hydrodynamics (GHD) has shed considerable light on their coarse-grained, large-scale properties~\cite{Doyon,  Fagotti, SciPostPhys.2.2.014, PhysRevLett.119.020602,  BBH0, BBH,PhysRevLett.119.020602, GHDII, doyon2017dynamics, solitongases,PhysRevLett.119.195301,2016arXiv160408434Z, PhysRevB.96.081118,PhysRevB.97.081111, dbd1, ghkv, dbd2, gv_superdiffusion, agrawal2019, gvw, horvath2019euler, PhysRevB.100.035108,2019arXiv190601654B}. The picture of dynamics that emerges from GHD (as well as complementary methods, such as exact bounds~\cite{PhysRevLett.82.1764, PhysRevLett.106.217206, PhysRevLett.111.057203,  mkp, carmelo2017, sirker:2010, 1742-5468-2014-9-P09037, urichuk2019spin} and large-scale numerics~\cite{lzp, PhysRevLett.122.210602, dupont_moore, weiner2019}) is rich and counterintuitive. Although a defining feature of integrable systems is the presence of stable, ballistically propagating quasiparticles, transport is not necessarily ballistic~\cite{PhysRevLett.78.943, PhysRevB.57.8307,PhysRevLett.95.187201}. Instead, some conserved quantities spread through regular or anomalous diffusion~\cite{idmp, dbd2, gv_superdiffusion, denardis_superdiffusion}, and even when ballistic transport is present, local autocorrelation functions  can decay with anomalous exponents~\cite{agrawal2019, gvw}. 

Perhaps the most surprising behavior, however, is that of the XXZ spin-$\frac{1}{2}$ chain with easy-plane anisotropy, governed by the Hamiltonian
\beq\label{xxzham}
H = J \sum\nolimits_i (S^x_i S^x_{i+1} + S^y_i S^y_{i+1} + \Delta S^z_i S^z_{i+1}).
\eeq
Here, $S_i^\alpha=\sigma^\alpha_i/2$  are spin-$\frac{1}{2}$ operators with $\sigma_i^\alpha$ the Pauli matrices on site $i$, the parameter $\Delta$ is the anisotropy, and $J$ is an overall coupling scale that we will set to unity in what follows. We consider $-1 < \Delta < 1$, so we can parameterize $\Delta \equiv \cos(\pi \lambda)$. For concreteness we assume that the system is at infinite temperature and in the thermodynamic limit. A remarkable feature of this model is as follows. 
Although spin transport is ballistic, the spin Drude weight $\mathcal{D}_\lambda$---defined below~\eqref{autoc}---appears to be discontinuous and fractal as a function of $\lambda$. Specifically, when $\lambda = p/q$ is rational, several distinct methods~\cite{PhysRevLett.82.1764, Prosen20141177,PhysRevB.96.081118,PhysRevLett.111.057203,PhysRevLett.119.020602, urichuk2019spin} lead to the conclusion that
\beq\label{drudeweight}
\mathcal{D}_\lambda = \frac{1}{12} (1 - \Delta^2) f\left(\frac{\pi}{q}\right), \quad f(x) = \frac{3}{2} \left[ \frac{1 - \frac{\sin(2 x)}{2x}} { \sin^2 x } \right].
\eeq
Eq.~\eqref{drudeweight} is known to be an exact lower bound on $\mathcal{D}$, which GHD~\cite{PhysRevLett.119.020602,BBH,PhysRevB.96.081118} predicts is saturated. 
Remarkably, Eq.~\eqref{drudeweight} allows the Drude weight to jump by ${\cal O}(1)$ as one changes $\Delta$ infinitesimally: $f(x) = 1$ for any irrational number, but is higher by an ${\cal O}(1)$ amount if there is a nearby rational number with a small denominator. 

\begin{figure}[!t]
\begin{center}
\includegraphics[width = 0.45\textwidth]{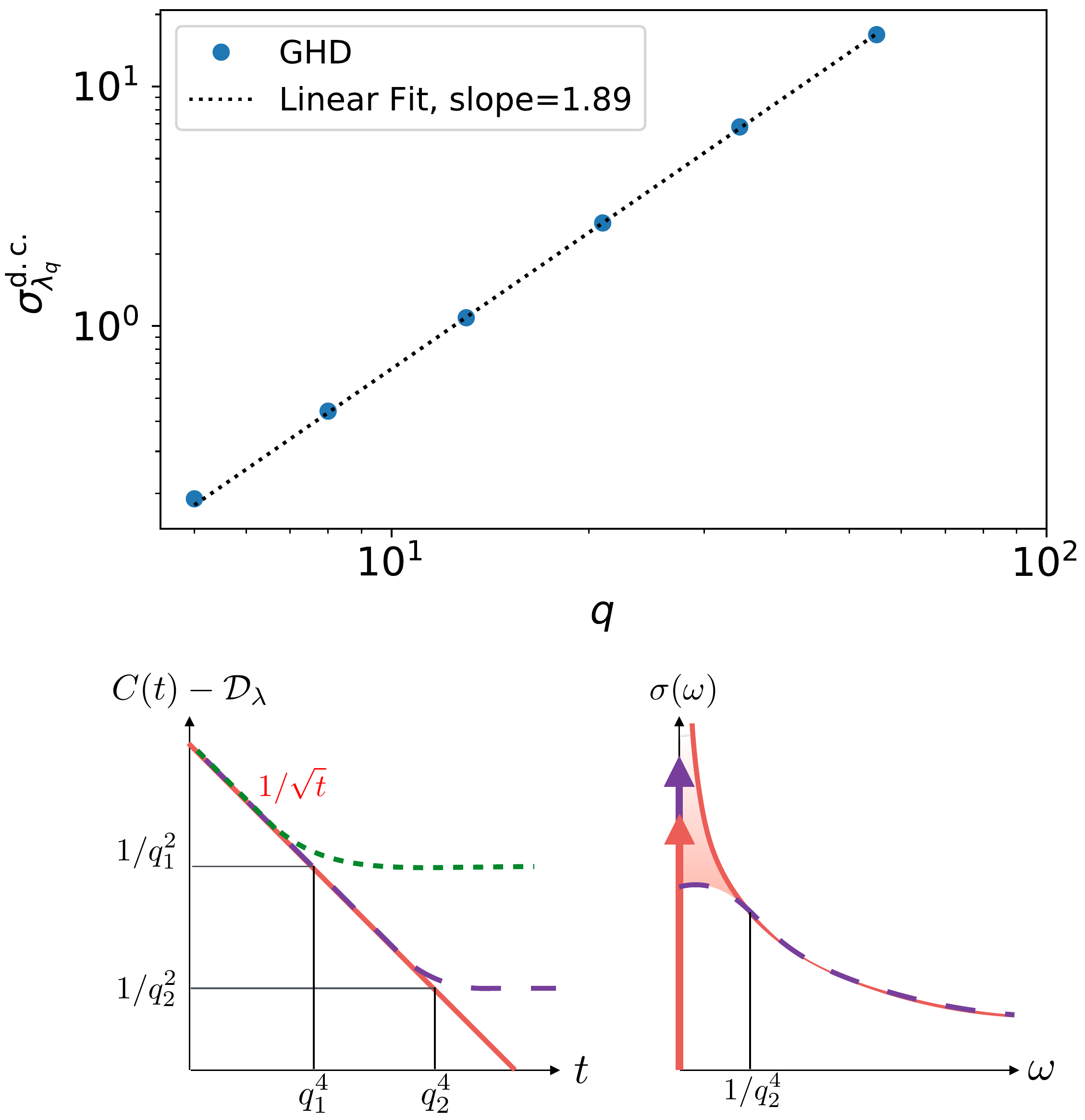}
\caption{Upper panel: value of the d.c. conductivity for rational Fibonacci approximants $\lambda_{q} = F_{n-1}/F_{n+1}$ {\it vs} $q = F_{n+1}$ to the generic irrational anisotropy $\lambda_\infty \equiv 1/\varphi^2$ where $\varphi$ is the Golden Ratio. We find $\sigma^{\rm d.c.}_{\lambda_q} \sim q^\beta$ with $\beta \approx 1.89$, corresponding to $\alpha \simeq 0.49$ in eq.~\eqref{basic_relation}. Lower panel: relationship between the d.c. conductivity for approximants, the crossover timescale, and the a.c. conductivity at the irrational $\lambda_\infty$. Left: the autocorrelation function $C(t)$ for $\lambda_\infty$ must follow that of a rational approximant with a given denominator $q_i$ until a crossover timescale $t^*_{q_i} \sim q_i^4$ (derived in the text). This forces $C(t) \sim 1/\sqrt{t}$ for $\lambda_\infty$. Right: in the frequency domain, the ``excess Drude weight'' at the rational approximant must precisely match the missing spectral weight in $\sigma(\omega)$ for $\omega < \omega^*_q \sim q^{-4}$.}
\label{fig1}
\end{center}
\end{figure}

It is physically clear that this rapid variation of the Drude weight must be accompanied by equally rapid variations of the low-frequency conductivity $\sigma(\omega)$, but the behavior of $\sigma(\omega)$ in this regime remains largely unexplored (but see Ref.~\cite{sanchez2018anomalous}), as it is not straightforward to compute in GHD. The present work combines GHD with exact constraints on the low-frequency behavior to arrive at a unified picture of transport in this unusual regime. We find that, for irrational $\lambda$, the quasiparticles responsible for spin transport undergo a L\'evy flight in addition to their ballistic motion; this leads to a low-frequency conductivity that scales as $\sigma(\omega) \sim 1/\sqrt{\omega}$. For rational $\lambda = p/q$, this behavior is cut off at frequencies $\omega^*_q \sim 1/q^4$, giving rise to a finite d.c. conductivity $\sim q^2$. Our strategy for deriving these results is to consider a series of rational approximants to a generic irrational $\lambda$; compute the d.c. conductivity for these using the recently discovered diffusive corrections to GHD~\cite{dbd1, dbd2, ghkv}; and relate $q$- and $\omega$-dependence through general constraints that follow from locality. Finally we compare our results with extensive simulations using matrix-product operators; the numerical results are consistent with a power-law divergence of $\sigma(\omega)$, although we cannot access late enough times to fix the exponent numerically. 

\emph{General constraints on $\sigma(\omega)$}.---The high-temperature limit of the conductivity $\sigma(\omega)$ is given by the Kubo formula
\beq\label{kubo}
\sigma_\lambda(\omega) \!=  \! \beta  \! \int_0^\infty \!\! dt \sum_x \!\! C_{jj}(x, t) e^{i \omega t} \! = \! \pi {\cal D}_\lambda \delta(\omega) \! + \! \sigma^{\rm reg}_\lambda(\omega) ,
\eeq
in terms of the autocorrelation function $C_{jj}(x,t)$ of the current $j(x) \equiv -i (S^+_x S^-_{x+1} - \mathrm{h.c.})$:
\beq\label{autoc}
C_{jj}(x,t; \lambda) \equiv Z^{-1} \mathrm{Tr}\left[ e^{i H_\lambda t} j(x) e^{-iH_\lambda t} j(0) e^{- \beta H_\lambda} \right].
\eeq
Here, $Z$ is the partition function and $\beta$ is the inverse temperature. 
In what follows we will suppress the subscript (since this paper treats only one correlation function) and also define $C_\lambda(t) \equiv \sum_x\, C_{jj} (x,t; \lambda)$. We will also take the $\beta \rightarrow 0$ limit. In this limit, all response functions including $\sigma(\omega)$ vanish, but the autocorrelation function~\eqref{autoc} is well-behaved, and therefore so is the quantity $\sigma(\omega)/\beta$~\eqref{kubo}: in the following, we will absorb this factor of $1/\beta$ in the definition of all transport coefficients. 
The Drude weight is defined as $\mathcal{D}_\lambda \equiv \lim_{t \rightarrow \infty} C_\lambda(t)$, and the d.c. conductivity is defined as $\sigma_{\lambda}^{\rm d.c.} \equiv \lim_{\omega \to 0} \sigma_{\lambda}^{\rm reg} (\omega)$. 

The autocorrelation function $C_\lambda(t)$, evaluated at any finite $t$, must be a continuous function of $\Delta$ and thus of $\lambda$: by the Lieb-Robinson theorem~\cite{Lieb2004}, one can truncate the infinite system on these timescales to a finite system of size $\propto t$, and all properties of finite systems must evolve continuously with $\Delta$. For some small $\varepsilon$, Eq.~\eqref{drudeweight} implies that one can find nearby values $\lambda, \lambda + \varepsilon$ such that $\mathcal{D}_\lambda$ and $\mathcal{D}_{\lambda + \varepsilon}$ differ by a large amount. Even so, locality implies that $|C_\lambda(t) - C_{\lambda + \varepsilon}(t)|$ remains small out to some late time $t^*$. One can easily show that $t^* \agt 1/\varepsilon$. In the frequency domain, the equivalent observation is that
\beq\label{weight1}
\int_0^\Omega d\omega |\sigma_\lambda(\omega) - \sigma_{\lambda+\epsilon}(\omega) | \alt C \frac{\varepsilon}{\Omega},
\eeq
where $\sigma_\lambda(\omega)$ is the \emph{full} conductivity~\eqref{kubo} at anisotropy $\lambda$, $\Omega > \varepsilon$ is generic, and $C$ is a constant of order unity. Changing $\lambda$ by $\varepsilon$ can only shift spectral weight over frequencies $\sim \varepsilon$. Thus there is a characteristic frequency $\omega^*(\varepsilon) \alt \varepsilon$ such that for $\omega \agt \omega^*(\varepsilon)$ the conductivity is essentially $\varepsilon$-independent. The drastic rearrangement of spectral weight that gives rise to the fractal structure of $\mathcal{D}_\lambda$~\eqref{drudeweight} must happen below this frequency (Fig.~\ref{fig1}). 

We now discuss how this constraint relates the a.c. conductivity of an irrational $\lambda$ to the d.c. conductivity of rational approximants. We approximate the irrational value, denoted $\lambda_\infty$ (and $\Delta_\infty = \cos \pi \lambda_\infty$), by a sequence of rational numbers $\{ \lambda_q = p/q \}$ with increasing denominators $q$. To simplify the discussion we will assume that $C_\lambda(t)$ decays monotonically at late times for all $\lambda$; within GHD this assumption certainly holds. By the reasoning above, until some late time $t^*_q$, $C_{\lambda_\infty}(t) \approx C_{ \lambda_q}(t) > \mathcal{D}_{\lambda_q}$. 
Assuming monotonicity, therefore, 
\beq
C_{ \lambda_\infty}(t) - \mathcal{D}_{\lambda_\infty} > \delta {\cal D}_{q}  \sim \frac{\pi^2 (1-\Delta_\infty)}{90 q^2},
\eeq
for all such large $q$, with $\delta {\cal D}_{q} \equiv {\cal D}_{\lambda_q}  - {\cal D}_{\lambda_\infty} $.

We consider the ansatz $C_{\lambda_\infty}(t) - \mathcal{D}_{\lambda_\infty} \sim 1/t^{1 - \alpha}$ [i.e., $\sigma_{\lambda_\infty}(\omega) \sim \omega^{-\alpha}$], for reasons that will become clear. This ansatz fixes the crossover timescale $t^*_q$ for large $q$, as follows. 
For $t \alt t^*_q$, $C_{\lambda_q}(t) \approx C_{ \lambda_\infty}(t)$, whereas for $t \agt t^*_q$, $C_{\lambda_q}(t) \approx \mathcal{D}_{\lambda_q}$. Equating the two forms at $t \sim t^*_q$ we find that $(t^*_q)^{1 - \alpha} \sim \delta {\cal D}^{-1}_q \sim q^2$, so
\beq\label{tq}
t^*_q \sim q^{2/(1 - \alpha)}.
\eeq
Finally, we relate this to the d.c. conductivity $\sigma_{\lambda_q}^{\rm d.c.}$ at  $\lambda_q$. This is the integral of $C_{\lambda_q}(t) - {\cal D}_{\lambda_q}$, which follows the power-law $1/t^{1-\alpha}$ and is cut off at time $t^*_q$. Combining this result with Eq.~\eqref{tq} we find that
\beq\label{basic_relation}
\sigma^{\rm d.c.}_{\lambda_q} \sim q^{2\alpha/(1 - \alpha)}, \quad \sigma_{\lambda_\infty}(\omega) \sim \omega^{-\alpha},
\eeq
where $\alpha \geq 0$. Note that this reasoning can be used to show that $\sigma(\omega)$ diverges, even without invoking GHD. By Dirichlet's approximation theorem, $|\lambda_q - \lambda_\infty| \alt 1/q^2$. Therefore, $t^*_q \agt q^2$, so $C_{\lambda_\infty}(t) - \mathcal{D}_{\lambda_\infty} \agt 1/t$. Fourier transforming gives $\sigma(\omega) \agt |\log \omega|$ at low frequencies, establishing that a divergence occurs. (This divergence had previously been predicted using GHD~\cite{idmp}.) 

Eq.~\eqref{basic_relation} can also be derived directly in frequency space, as follows. Suppose $\sigma_{\lambda_\infty}(\omega) \sim \omega^{-\alpha}$. Then by Eq.~\eqref{weight1}, $\int_{0}^{\omega^*_q} d\omega [\sigma^{\rm reg}_{\lambda_\infty}(\omega) - \sigma^{\rm d.c.}_{\lambda_q}] \simeq \delta {\cal D}_q \sim 1/q^2$: the extra Drude weight at the commensurate point must precisely match the missing part of the regular spectral weight (Fig.~\ref{fig1}).
Thus, $[\omega^*_q]^{1-\alpha} \sim 1/q^2$, so $\omega^*_q \sim q^{-2/(1-\alpha)}$, consistent with Eq.~\eqref{tq} and $\omega^*_q \sim 1/t^*_q$. 
This relation between the exponents governing the frequency-dependence of the conductivity and the $q$-dependence of the d.c. conductivity for approximants will be crucial in what follows. 

\begin{figure*}[tb]
\begin{center}
\includegraphics[width = 0.98 \textwidth]{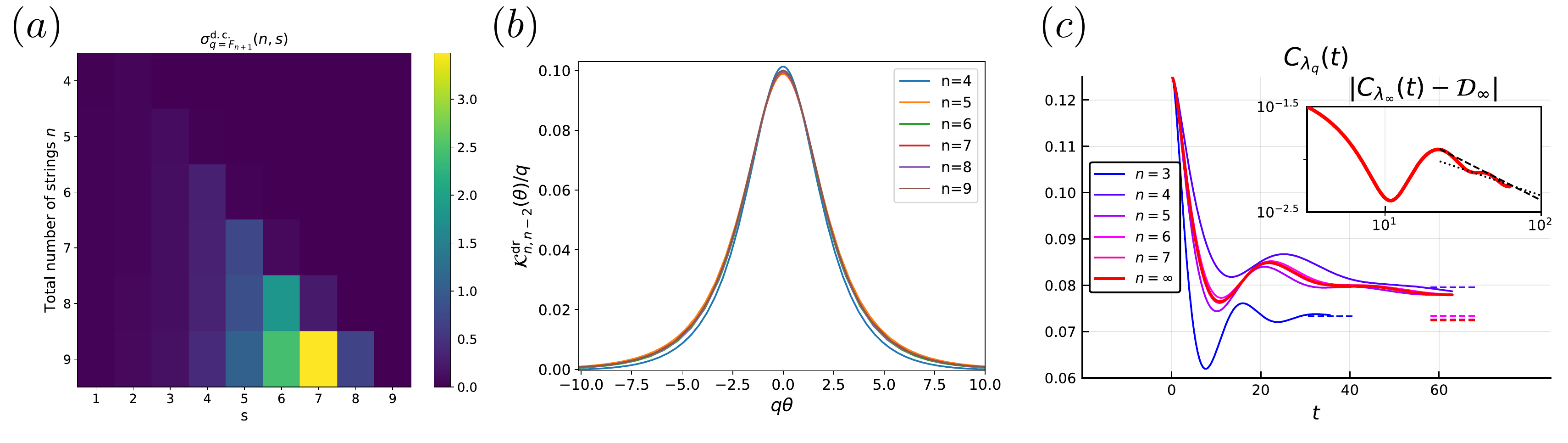}
\caption{(a)~Contributions to the d.c. conductivity of the charged quasiparticle from scattering off each species of neutral quasiparticle. For any given $n$, the dominant source of diffusion is the heaviest neutral string $n - 2$. (b)~Rapidity-dependence of the dressed kernel for scattering between the charged quasiparticle and the largest neutral quasiparticle; we find that $K^{\mathrm{dr}}_{n, n-2} (\theta)$ has a peak of height $q_n$ and width $1/q_n$, as shown by the data collapse. (c)~TEBD numerics for the current-current correlator for various $n$; plots for the larger $n$ stay close to the $n = \infty$ value at the accessible times. Inset: Power-law decay of $C_{\lambda_\infty} (t)- {\cal D}_{\lambda_\infty}$: although our time range is limited, our data is consistent with an exponent $1-\alpha \in (\frac{1}{2}, \frac{3}{4})$ (dashed lines).
}
\label{bigfig}
\end{center}
\end{figure*}

\emph{Generalized hydrodynamics}.---The argument above shows that very generally $\sigma(\omega)$ must diverge at low frequencies for irrational values of $\lambda$. However, on its own it is insufficient to determine the exponent $\alpha$. To do this we adopt the framework of generalized hydrodynamics (GHD)~\cite{Fagotti,Doyon}, which was recently extended to incorporate diffusion~\cite{dbd1, dbd2, ghkv}. 
GHD allows one to compute the response of an integrable system in the hydrodynamic regime, i.e., when one can take the system to be locally in equilibrium in a generalized Gibbs ensemble~\cite{PhysRevLett.98.050405, caux_konik, 1742-5468-2013-07-P07012, 1742-5468-2013-07-P07003, PhysRevLett.113.117203, PhysRevLett.115.157201, PhysRevLett.113.117202, 1742-5468-2016-6-064007, Pozsgay}. 
In this regime, the dynamics of an integrable system maps onto that of an appropriate classical soliton gas~\cite{solitongases}. Solitons (i.e., the quasiparticles of the integrable system) propagate ballistically, but acquire time-delays when they scatter elastically off each other~\cite{solitongases, el2003thermodynamic, sg_ffa, ghkv}. Gaussian fluctuations of the quasiparticle densities lead to fluctuations of the distance traveled by each quasiparticle~\cite{ghkv}. Each quasiparticle can thus be understood as following a biased random walk. Since the quasiparticles carry conserved charges, such as spin, the conserved charges also pick up a sub-leading diffusive correction to their ballistic transport: the variance of the spin current is directly related to the variance of the quasiparticle velocities due to collisions. 
For the spin d.c. conductivity $\sigma_{\lambda_q}^{\rm d.c.}$, we have the relation~\cite{dbd1,dbd2}
\bea\label{dc}
\sigma_{\lambda_q}^{\rm d.c.} & = & \! \frac{1}{4} \!\sum_{kl} \! \int \! \! d \theta_1 d\theta_2 \rho_k(\theta_1\!) \rho_l(\theta_2\!) f_k f_l  |v_k(\theta_1\!)\! -\! v_l (\theta_2\!)| \nonumber \\
 & & \, \times \! \!\left[ {\cal K}_{kl}^{\mathrm{dr}} (\theta_1-\theta_2)\!\! \left(\!\frac{m^{\mathrm{dr}}_k  }{\rho^{\rm tot}_k (\theta_1\!) \sigma_k} \! - \! \frac{m^{\mathrm{dr}}_l  }{\rho^{\rm tot}_l (\theta_2\!) \sigma_l} \!\right)\!\right]^2\!\!\!,
\eea
in terms of data from the Thermodynamic Bethe Ansatz (TBA)~\cite{takahashi}. 
In this expression, $k,l$ label quasiparticle species and $\theta_i$ label rapidities; and the other symbols denote properties (within the TBA) of quasiparticles with labels $k, \theta$: 
$\rho_k(\theta)$ is the density of quasiparticles; $f_k = 1 - \rho_k(\theta)/\rho^{\mathrm{tot}}_k(\theta)$ is related to their filling factor (independent from $\theta$ at infinite temperature); $\rho^{\mathrm{tot}}_k(\theta)$ is the total density of states; $v_k(\theta)$ and $m^{\mathrm{dr}}_k$ are respectively the dressed velocity---derived from the dressed dispersion relation---and dressed magnetization; and $\sigma_k = \pm 1$ is the so-called $\sigma$-\emph{parity} of quasiparticle species $k$. The dressed kernel $\mathcal{K}^{\mathrm{dr}}$ is the solution to an integral equation that has to be solved numerically~\cite{suppmat}. A brief overview of the TBA formalism as it applies here is given in~\cite{suppmat}; for more details we refer to Ref.~\cite{takahashi}.

As a generic irrational number, we choose $\lambda_\infty = 1 / \varphi^2$ where $\varphi$ is the Golden Ratio. This number is generic in the sense that it is poorly approximable by rational numbers; in this sense it resembles almost all real numbers~\cite{bernik2013distribution}. Moreover, the TBA for this number has the advantage of being tractable, with a simple quasiparticle hierarchy.
We will discuss exceptional cases briefly at the end. The continued fraction expansion of $\varphi^2 = 1/(2 + 1/(1 + \ldots))$. One can truncate this expansion by replacing the last term with $2$; this gives the series $\lambda_n =  F_{n-1} / F_{n+1}$, where $F_n$ is the $n$th Fibonacci number. The Bethe ansatz solution for $\lambda_n$ involves $n $ quasiparticle species. At zero field, the first $n-2$ quasiparticle species carry no dressed magnetization; the last two quasiparticle species each carry a magnetization $\sim F_{n+1} = q$ and are responsible for the finite spin Drude weight. We will refer to quasiparticles with larger values of $n$ as being ``larger,'' which is true at the lattice scale; however, within GHD one treats all quasiparticles as pointlike. Spin transport is dominated by charged quasiparticles; the other, ``neutral'' quasiparticles affect spin transport by scattering elastically off the charged quasiparticles and causing them to diffuse. 

GHD yields the following conclusions for spin transport. Charged quasiparticles move with a characteristic velocity which saturates to an ${\cal O}(1)$ value as $q \to \infty$, and as they move they scatter off neutral strings. Large neutral strings are rare $\rho_{n-2} \sim q^{-2}$, but also have an outsized influence, because their scattering phase shifts are large. 
Fig.~\ref{bigfig} separates out the contributions to the spin diffusion constant by quasiparticle index/size: it is clear that the dominant contribution comes from scattering off the largest neutral quasiparticle. 
Explicitly evaluating Eq.~\eqref{dc} with the appropriate TBA data we find that $\sigma_{\mathrm{d.c.}}(\lambda_q) \sim q^2$. This result can be derived analytically~\cite{suppmat}, and is consistent with numerical evaluation of Eq.~\eqref{dc} [Fig.~\ref{fig1}]. Using Eq.~\eqref{basic_relation} this means that 
\bea
\sigma_{\lambda_\infty}(\omega) \sim 1/\sqrt{\omega}, 
\eea
and therefore that $t^*_q \sim q^4$. 

\emph{Soliton gas picture}.---This long crossover timescale has a physical interpretation in terms of the semiclassical soliton gas framework~\cite{solitongases, ghkv}. The dressed kernel ${\cal K}^{\mathrm{dr}}(\theta)$ is peaked at $\theta = 0$, with a peak height that scales as $q$ and a peak width that scales as $1/q$ (Fig.~\ref{bigfig}). The dominant scattering events that a charged particle experiences are those with large neutral quasiparticles which have almost the same rapidity and therefore almost the same velocity (up to $\sim 1/q$). At large $q$ the heaviest neutral quasiparticle has density $1/q^2$; fixing its rapidity to a window of size $1/q$ reduces the density of dominant scatterers to $1/q^3$. Since the two quasiparticles start out spaced at a distance $q^3$ and have a relative velocity $\sim 1/q$, they collide on a timescale $t_q^* \sim q^4$. At much shorter timescales, the system is not in local equilibrium and the asymptotic result~\eqref{dc} does not apply. 

One can derive further physical insight by applying the soliton-gas framework to the motion of the charged quasiparticle at very large $q$ but for $t \ll t_q^*$. In this regime, as time passes, the charged quasiparticle encounters increasingly large neutral quasiparticles, and therefore picks up increasingly large displacements. For instance, in a time $t$, the largest collision event will involve a string for which $q(t) \sim t^{1/4}$. This string gives a (dressed) displacement~\cite{solitongases,BBH,2019arXiv190702474V} of order $\Delta x^{\rm dr} = {\cal K}^{\rm dr}/p'(\theta) \sim q^3 \sim t^{3/4}$ where we have used the fact that the dressed momentum scales as $p'(\theta) \sim \rho^{\rm tot}(\theta)\sim q^{-2}$. Therefore, the variance of the position of the charged string scales as $t^{3/2}$, consistent with our exponent for the conductivity. Since the dynamics of the charged quasiparticle consists of kicks of power-law increasing strength, whose probability also falls off as a power law, it is a canonical example of a L\'evy flight, with the parameter $\mu = 4/3$ in the notation of Ref.~\cite{bouchaud_georges}. An interesting direction for future work would be to compare the spin structure factor to the scaling forms predicted for L\'evy flights in that work.

\emph{TEBD simulations}.---To check our assumptions and the GHD analysis, we have also performed explicit numerical calculations of the correlation function $C(t)$~\eqref{autoc}. This calculation can be simplified by exploiting translation-invariance. At infinite temperature Eq.~\eqref{autoc} takes the form $C(x,t) = 2^{-L} \mathrm{Tr}[j(x,t/2) j(0, -t/2)]$. By translation-invariance, the operator $j(x,t/2)$ is just a translated version of the operator $j(0,t/2)$ --- which in turn is minus the conjugate of $j(0,-t/2)$. Thus to evaluate the full autocorrelator~\eqref{autoc} it suffices to evaluate the time evolution of a \emph{single} local operator. This can be done using the time-evolving block decimation (TEBD) method~\cite{PhysRevLett.91.147902,vidal,whitetdmrg} for matrix-product operators (MPOs)~\cite{vgrc,schollwoeck}. Exploiting this simplification allows us to save considerable computational overhead, and study systems in the thermodynamic limit for times up to $t \approx 60$. To reach such long times, we fix the bond dimension to $\chi=512$ instead of the truncation error. At even longer times, errors accumulate and give clearly unphysical results~\cite{suppmat}, but at the times for which we have shown data (Fig.~\ref{bigfig}) the errors remain small. 

Our results are shown in Fig.~\ref{bigfig}. At the accessible timescales, data for the larger $n$ stay close to that for $n = 4$ and far from their asymptotic values (dashed lines). This confirms our picture that the decay is a slow process. The inset shows the evolution of $C_{\lambda_\infty}(t) - \mathcal{D}_\lambda$ on a log-log plot; although our dynamic range is limited, the data support a power law, with an exponent $1-\alpha \in (\frac{1}{2}, \frac{3}{4})$, roughly consistent with our predictions. Note that at times $t \sim 100$ one can only reach the crossover timescale for $q \sim 3$, so we are far from the asymptotic behavior. 

\emph{Isotropic limit}.--- In closing, we note that our approach can also be applied to approach the isotropic point $\Delta =1$ by using the rational series $\lambda_q=1/q$ as $q \to \infty$. In that limit, the Drude weight vanishes ${\cal D}_{\lambda_\infty} = 0$, but we still have $\delta {\cal D}_q \sim q^{-2}$ from~\eqref{drudeweight}. Therefore, equations~\eqref{basic_relation} also hold in this case, but eq.~\eqref{dc} now predicts $\sigma_q^{\rm d.c.} \sim q$, corresponding to $\alpha=1/3$. This immediately leads to the prediction of spin superdiffusion $\sigma_{\lambda_\infty}(\omega) \sim \omega^{-1/3}$ at the isotopic point, consistent with earlier results~\cite{PhysRevLett.106.220601,lzp,gv_superdiffusion}. Remarkably, the corresponding time scale $t_q^* \sim q^3$ was also identified approaching the isotropic point from $\Delta >1$ in Ref.~\cite{gv_superdiffusion}. Importantly, the L\'evy flight mechanism discussed in the easy-plane case is different from the (conjectured) Kardar-Parisi-Zhang mechanism for superdiffusion in the $\Delta = 1$ case~\cite{PhysRevLett.122.210602,gvw,denardis_superdiffusion}, in terms of both the dynamical exponents and the expected shape of the front. Our results therefore suggest that anomalous transport signatures are generic in integrable systems with an infinite quasiparticle hierarchy. This naturally raises the question of whether there are ``universality classes'' of integrable dynamics, and of what types of dynamical scaling are possible in integrable models.

\emph{Discussion}.---In this work we studied the finite-frequency response of the XXZ model in its easy-plane (``gapless'') phase, both numerically and within the GHD framework. We showed that this response is generically anomalous, and is described by a L\'evy flight of the charged quasiparticles due to collisions with neutral quasiparticles. Using a combination of GHD and exact arguments, we showed that the optical conductivity in this regime generically diverges as $\sigma(\omega) \sim 1/\sqrt{\omega}$, for a certain class of irrational $\lambda$, i.e., those that are ``farthest'' from rational. For strictly rational $\lambda$, $\sigma(\omega)$ is cut off at low frequencies and the low-frequency spectral weight migrates to the Drude peak. An interesting question is what happens for irrational numbers with more unbounded continued fraction expansions; qualitatively, such irrationals are parametrically ``closer'' to rational approximants, so we expect longer crossover timescales and thus even steeper divergences in $\sigma(\omega)$ for this case. In the specific instance of Liouville numbers, adapting our continuity arguments leads to the conclusion that $\sigma(\omega) \sim 1/\omega$ up to sub-power-law corrections (since by definition of such numbers $|\lambda - \lambda_q| < 1/q^n$ for any $n$, implying that $t^*_q > q^n$ for any $n$). Testing this within GHD is an interesting topic for future work. 

Our results explain how to account for the ``excess'' Drude weight at rational $\lambda$: this Drude weight is drawn from the low-frequency conductivity (Fig.~\ref{fig1}). Our work both resolves many of the outstanding puzzles about this regime and sheds light on features that had not yet been predicted, such as the low-frequency divergence of $\sigma(\omega)$ and the prediction that the spin front should match that of a L\'evy flight. These predictions are an important question for future numerical work to investigate.

\begin{acknowledgments}

\emph{Acknowledgments}.---The authors thank Jacopo De Nardis, David Huse, and Vadim Oganesyan for useful discussions, and Jacopo De Nardis and Toma\v{z} Prosen for helpful comments on the manuscript. This work was supported by the National Science Foundation under NSF Grant No. DMR-1653271 (S.G.),  the US Department of Energy, Office of Science, Basic Energy Sciences, under Early Career Award No. DE-SC0019168 (U.A. and R.V.),  and the Alfred P. Sloan Foundation through a Sloan Research Fellowship (R.V.). 

\end{acknowledgments}

\bibliography{refs}

\begin{thebibliography}{77}%
\makeatletter
\providecommand \@ifxundefined [1]{%
 \@ifx{#1\undefined}
}%
\providecommand \@ifnum [1]{%
 \ifnum #1\expandafter \@firstoftwo
 \else \expandafter \@secondoftwo
 \fi
}%
\providecommand \@ifx [1]{%
 \ifx #1\expandafter \@firstoftwo
 \else \expandafter \@secondoftwo
 \fi
}%
\providecommand \natexlab [1]{#1}%
\providecommand \enquote  [1]{``#1''}%
\providecommand \bibnamefont  [1]{#1}%
\providecommand \bibfnamefont [1]{#1}%
\providecommand \citenamefont [1]{#1}%
\providecommand \href@noop [0]{\@secondoftwo}%
\providecommand \href [0]{\begingroup \@sanitize@url \@href}%
\providecommand \@href[1]{\@@startlink{#1}\@@href}%
\providecommand \@@href[1]{\endgroup#1\@@endlink}%
\providecommand \@sanitize@url [0]{\catcode `\\12\catcode `\$12\catcode
  `\&12\catcode `\#12\catcode `\^12\catcode `\_12\catcode `\%12\relax}%
\providecommand \@@startlink[1]{}%
\providecommand \@@endlink[0]{}%
\providecommand \url  [0]{\begingroup\@sanitize@url \@url }%
\providecommand \@url [1]{\endgroup\@href {#1}{\urlprefix }}%
\providecommand \urlprefix  [0]{URL }%
\providecommand \Eprint [0]{\href }%
\providecommand \doibase [0]{http://dx.doi.org/}%
\providecommand \selectlanguage [0]{\@gobble}%
\providecommand \bibinfo  [0]{\@secondoftwo}%
\providecommand \bibfield  [0]{\@secondoftwo}%
\providecommand \translation [1]{[#1]}%
\providecommand \BibitemOpen [0]{}%
\providecommand \bibitemStop [0]{}%
\providecommand \bibitemNoStop [0]{.\EOS\space}%
\providecommand \EOS [0]{\spacefactor3000\relax}%
\providecommand \BibitemShut  [1]{\csname bibitem#1\endcsname}%
\let\auto@bib@innerbib\@empty
\bibitem [{\citenamefont {Prosen}(2011)}]{PhysRevLett.106.217206}%
  \BibitemOpen
  \bibfield  {author} {\bibinfo {author} {\bibfnamefont {T.}~\bibnamefont
  {Prosen}},\ }\href {\doibase 10.1103/PhysRevLett.106.217206} {\bibfield
  {journal} {\bibinfo  {journal} {Phys. Rev. Lett.}\ }\textbf {\bibinfo
  {volume} {106}},\ \bibinfo {pages} {217206} (\bibinfo {year}
  {2011})}\BibitemShut {NoStop}%
\bibitem [{\citenamefont {Caux}\ and\ \citenamefont
  {Essler}(2013)}]{PhysRevLett.110.257203}%
  \BibitemOpen
  \bibfield  {author} {\bibinfo {author} {\bibfnamefont {J.-S.}\ \bibnamefont
  {Caux}}\ and\ \bibinfo {author} {\bibfnamefont {F.~H.~L.}\ \bibnamefont
  {Essler}},\ }\href {\doibase 10.1103/PhysRevLett.110.257203} {\bibfield
  {journal} {\bibinfo  {journal} {Phys. Rev. Lett.}\ }\textbf {\bibinfo
  {volume} {110}},\ \bibinfo {pages} {257203} (\bibinfo {year}
  {2013})}\BibitemShut {NoStop}%
\bibitem [{\citenamefont {Wouters}\ \emph {et~al.}(2014)\citenamefont
  {Wouters}, \citenamefont {De~Nardis}, \citenamefont {Brockmann},
  \citenamefont {Fioretto}, \citenamefont {Rigol},\ and\ \citenamefont
  {Caux}}]{PhysRevLett.113.117202}%
  \BibitemOpen
  \bibfield  {author} {\bibinfo {author} {\bibfnamefont {B.}~\bibnamefont
  {Wouters}}, \bibinfo {author} {\bibfnamefont {J.}~\bibnamefont {De~Nardis}},
  \bibinfo {author} {\bibfnamefont {M.}~\bibnamefont {Brockmann}}, \bibinfo
  {author} {\bibfnamefont {D.}~\bibnamefont {Fioretto}}, \bibinfo {author}
  {\bibfnamefont {M.}~\bibnamefont {Rigol}}, \ and\ \bibinfo {author}
  {\bibfnamefont {J.-S.}\ \bibnamefont {Caux}},\ }\href {\doibase
  10.1103/PhysRevLett.113.117202} {\bibfield  {journal} {\bibinfo  {journal}
  {Phys. Rev. Lett.}\ }\textbf {\bibinfo {volume} {113}},\ \bibinfo {pages}
  {117202} (\bibinfo {year} {2014})}\BibitemShut {NoStop}%
\bibitem [{\citenamefont {Ilievski}\ \emph {et~al.}(2015)\citenamefont
  {Ilievski}, \citenamefont {De~Nardis}, \citenamefont {Wouters}, \citenamefont
  {Caux}, \citenamefont {Essler},\ and\ \citenamefont
  {Prosen}}]{PhysRevLett.115.157201}%
  \BibitemOpen
  \bibfield  {author} {\bibinfo {author} {\bibfnamefont {E.}~\bibnamefont
  {Ilievski}}, \bibinfo {author} {\bibfnamefont {J.}~\bibnamefont {De~Nardis}},
  \bibinfo {author} {\bibfnamefont {B.}~\bibnamefont {Wouters}}, \bibinfo
  {author} {\bibfnamefont {J.-S.}\ \bibnamefont {Caux}}, \bibinfo {author}
  {\bibfnamefont {F.~H.~L.}\ \bibnamefont {Essler}}, \ and\ \bibinfo {author}
  {\bibfnamefont {T.}~\bibnamefont {Prosen}},\ }\href {\doibase
  10.1103/PhysRevLett.115.157201} {\bibfield  {journal} {\bibinfo  {journal}
  {Phys. Rev. Lett.}\ }\textbf {\bibinfo {volume} {115}},\ \bibinfo {pages}
  {157201} (\bibinfo {year} {2015})}\BibitemShut {NoStop}%
\bibitem [{\citenamefont {Takahashi}(1999)}]{takahashi}%
  \BibitemOpen
  \bibfield  {author} {\bibinfo {author} {\bibfnamefont {M.}~\bibnamefont
  {Takahashi}},\ }\href {https://books.google.com/books?id=kX1FAwEACAAJ} {\emph
  {\bibinfo {title} {Thermodynamics of One-Dimensional Solvable Models}}}\
  (\bibinfo  {publisher} {Cambridge University Press},\ \bibinfo {year}
  {1999})\BibitemShut {NoStop}%
\bibitem [{\citenamefont {Calabrese}\ and\ \citenamefont
  {Cardy}(2006)}]{Calabrese:2006}%
  \BibitemOpen
  \bibfield  {author} {\bibinfo {author} {\bibfnamefont {P.}~\bibnamefont
  {Calabrese}}\ and\ \bibinfo {author} {\bibfnamefont {J.}~\bibnamefont
  {Cardy}},\ }\href@noop {} {\bibfield  {journal} {\bibinfo  {journal}
  {Physical Review Letters}\ }\textbf {\bibinfo {volume} {96}},\ \bibinfo
  {pages} {136801} (\bibinfo {year} {2006})}\BibitemShut {NoStop}%
\bibitem [{\citenamefont {Pozsgay}\ \emph {et~al.}(2014)\citenamefont
  {Pozsgay}, \citenamefont {Mesty\'an}, \citenamefont {Werner}, \citenamefont
  {Kormos}, \citenamefont {Zar\'and},\ and\ \citenamefont
  {Tak\'acs}}]{PhysRevLett.113.117203}%
  \BibitemOpen
  \bibfield  {author} {\bibinfo {author} {\bibfnamefont {B.}~\bibnamefont
  {Pozsgay}}, \bibinfo {author} {\bibfnamefont {M.}~\bibnamefont {Mesty\'an}},
  \bibinfo {author} {\bibfnamefont {M.~A.}\ \bibnamefont {Werner}}, \bibinfo
  {author} {\bibfnamefont {M.}~\bibnamefont {Kormos}}, \bibinfo {author}
  {\bibfnamefont {G.}~\bibnamefont {Zar\'and}}, \ and\ \bibinfo {author}
  {\bibfnamefont {G.}~\bibnamefont {Tak\'acs}},\ }\href {\doibase
  10.1103/PhysRevLett.113.117203} {\bibfield  {journal} {\bibinfo  {journal}
  {Phys. Rev. Lett.}\ }\textbf {\bibinfo {volume} {113}},\ \bibinfo {pages}
  {117203} (\bibinfo {year} {2014})}\BibitemShut {NoStop}%
\bibitem [{\citenamefont {Ilievski}\ \emph {et~al.}(2016)\citenamefont
  {Ilievski}, \citenamefont {Medenjak}, \citenamefont {Prosen},\ and\
  \citenamefont {Zadnik}}]{2016arXiv160300440I}%
  \BibitemOpen
  \bibfield  {author} {\bibinfo {author} {\bibfnamefont {E.}~\bibnamefont
  {Ilievski}}, \bibinfo {author} {\bibfnamefont {M.}~\bibnamefont {Medenjak}},
  \bibinfo {author} {\bibfnamefont {T.}~\bibnamefont {Prosen}}, \ and\ \bibinfo
  {author} {\bibfnamefont {L.}~\bibnamefont {Zadnik}},\ }\href
  {http://stacks.iop.org/1742-5468/2016/i=6/a=064008} {\bibfield  {journal}
  {\bibinfo  {journal} {Journal of Statistical Mechanics: Theory and
  Experiment}\ }\textbf {\bibinfo {volume} {2016}},\ \bibinfo {pages} {064008}
  (\bibinfo {year} {2016})}\BibitemShut {NoStop}%
\bibitem [{\citenamefont {Calabrese}\ \emph {et~al.}(2016)\citenamefont
  {Calabrese}, \citenamefont {Essler},\ and\ \citenamefont
  {Mussardo}}]{1742-5468-2016-6-064001}%
  \BibitemOpen
  \bibfield  {author} {\bibinfo {author} {\bibfnamefont {P.}~\bibnamefont
  {Calabrese}}, \bibinfo {author} {\bibfnamefont {F.~H.~L.}\ \bibnamefont
  {Essler}}, \ and\ \bibinfo {author} {\bibfnamefont {G.}~\bibnamefont
  {Mussardo}},\ }\href {http://stacks.iop.org/1742-5468/2016/i=6/a=064001}
  {\bibfield  {journal} {\bibinfo  {journal} {Journal of Statistical Mechanics:
  Theory and Experiment}\ }\textbf {\bibinfo {volume} {2016}},\ \bibinfo
  {pages} {064001} (\bibinfo {year} {2016})}\BibitemShut {NoStop}%
\bibitem [{\citenamefont {Essler}\ and\ \citenamefont
  {Fagotti}(2016)}]{1742-5468-2016-6-064002}%
  \BibitemOpen
  \bibfield  {author} {\bibinfo {author} {\bibfnamefont {F.~H.~L.}\
  \bibnamefont {Essler}}\ and\ \bibinfo {author} {\bibfnamefont
  {M.}~\bibnamefont {Fagotti}},\ }\href
  {http://stacks.iop.org/1742-5468/2016/i=6/a=064002} {\bibfield  {journal}
  {\bibinfo  {journal} {Journal of Statistical Mechanics: Theory and
  Experiment}\ }\textbf {\bibinfo {volume} {2016}},\ \bibinfo {pages} {064002}
  (\bibinfo {year} {2016})}\BibitemShut {NoStop}%
\bibitem [{\citenamefont {Vasseur}\ and\ \citenamefont
  {Moore}(2016)}]{1742-5468-2016-6-064010}%
  \BibitemOpen
  \bibfield  {author} {\bibinfo {author} {\bibfnamefont {R.}~\bibnamefont
  {Vasseur}}\ and\ \bibinfo {author} {\bibfnamefont {J.~E.}\ \bibnamefont
  {Moore}},\ }\href {http://stacks.iop.org/1742-5468/2016/i=6/a=064010}
  {\bibfield  {journal} {\bibinfo  {journal} {Journal of Statistical Mechanics:
  Theory and Experiment}\ }\textbf {\bibinfo {volume} {2016}},\ \bibinfo
  {pages} {064010} (\bibinfo {year} {2016})}\BibitemShut {NoStop}%
\bibitem [{\citenamefont {Vidmar}\ and\ \citenamefont
  {Rigol}(2016)}]{1742-5468-2016-6-064007}%
  \BibitemOpen
  \bibfield  {author} {\bibinfo {author} {\bibfnamefont {L.}~\bibnamefont
  {Vidmar}}\ and\ \bibinfo {author} {\bibfnamefont {M.}~\bibnamefont {Rigol}},\
  }\href {http://stacks.iop.org/1742-5468/2016/i=6/a=064007} {\bibfield
  {journal} {\bibinfo  {journal} {Journal of Statistical Mechanics: Theory and
  Experiment}\ }\textbf {\bibinfo {volume} {2016}},\ \bibinfo {pages} {064007}
  (\bibinfo {year} {2016})}\BibitemShut {NoStop}%
\bibitem [{\citenamefont {Fagotti}\ \emph {et~al.}(2014)\citenamefont
  {Fagotti}, \citenamefont {Collura}, \citenamefont {Essler},\ and\
  \citenamefont {Calabrese}}]{PhysRevB.89.125101}%
  \BibitemOpen
  \bibfield  {author} {\bibinfo {author} {\bibfnamefont {M.}~\bibnamefont
  {Fagotti}}, \bibinfo {author} {\bibfnamefont {M.}~\bibnamefont {Collura}},
  \bibinfo {author} {\bibfnamefont {F.~H.~L.}\ \bibnamefont {Essler}}, \ and\
  \bibinfo {author} {\bibfnamefont {P.}~\bibnamefont {Calabrese}},\ }\href
  {\doibase 10.1103/PhysRevB.89.125101} {\bibfield  {journal} {\bibinfo
  {journal} {Phys. Rev. B}\ }\textbf {\bibinfo {volume} {89}},\ \bibinfo
  {pages} {125101} (\bibinfo {year} {2014})}\BibitemShut {NoStop}%
\bibitem [{\citenamefont {Alba}\ and\ \citenamefont
  {Calabrese}(2017)}]{alba2017entanglement}%
  \BibitemOpen
  \bibfield  {author} {\bibinfo {author} {\bibfnamefont {V.}~\bibnamefont
  {Alba}}\ and\ \bibinfo {author} {\bibfnamefont {P.}~\bibnamefont
  {Calabrese}},\ }\href@noop {} {\bibfield  {journal} {\bibinfo  {journal}
  {Proceedings of the National Academy of Sciences}\ }\textbf {\bibinfo
  {volume} {114}},\ \bibinfo {pages} {7947} (\bibinfo {year}
  {2017})}\BibitemShut {NoStop}%
\bibitem [{\citenamefont {Kinoshita}\ \emph {et~al.}(2006)\citenamefont
  {Kinoshita}, \citenamefont {Wenger},\ and\ \citenamefont
  {Weiss}}]{kinoshita}%
  \BibitemOpen
  \bibfield  {author} {\bibinfo {author} {\bibfnamefont {T.}~\bibnamefont
  {Kinoshita}}, \bibinfo {author} {\bibfnamefont {T.}~\bibnamefont {Wenger}}, \
  and\ \bibinfo {author} {\bibfnamefont {D.}~\bibnamefont {Weiss}},\
  }\href@noop {} {\bibfield  {journal} {\bibinfo  {journal} {Nature}\ }\textbf
  {\bibinfo {volume} {440}},\ \bibinfo {pages} {900} (\bibinfo {year}
  {2006})}\BibitemShut {NoStop}%
\bibitem [{\citenamefont {Gring}\ \emph {et~al.}(2012)\citenamefont {Gring},
  \citenamefont {Kuhnert}, \citenamefont {Langen}, \citenamefont {Kitagawa},
  \citenamefont {Rauer}, \citenamefont {Schreitl}, \citenamefont {Mazets},
  \citenamefont {Smith}, \citenamefont {Demler},\ and\ \citenamefont
  {Schmiedmayer}}]{gring}%
  \BibitemOpen
  \bibfield  {author} {\bibinfo {author} {\bibfnamefont {M.}~\bibnamefont
  {Gring}}, \bibinfo {author} {\bibfnamefont {M.}~\bibnamefont {Kuhnert}},
  \bibinfo {author} {\bibfnamefont {T.}~\bibnamefont {Langen}}, \bibinfo
  {author} {\bibfnamefont {T.}~\bibnamefont {Kitagawa}}, \bibinfo {author}
  {\bibfnamefont {B.}~\bibnamefont {Rauer}}, \bibinfo {author} {\bibfnamefont
  {M.}~\bibnamefont {Schreitl}}, \bibinfo {author} {\bibfnamefont
  {I.}~\bibnamefont {Mazets}}, \bibinfo {author} {\bibfnamefont {D.~A.}\
  \bibnamefont {Smith}}, \bibinfo {author} {\bibfnamefont {E.}~\bibnamefont
  {Demler}}, \ and\ \bibinfo {author} {\bibfnamefont {J.}~\bibnamefont
  {Schmiedmayer}},\ }\href {\doibase 10.1126/science.1224953} {\bibfield
  {journal} {\bibinfo  {journal} {Science}\ }\textbf {\bibinfo {volume}
  {337}},\ \bibinfo {pages} {1318} (\bibinfo {year} {2012})}\BibitemShut
  {NoStop}%
\bibitem [{\citenamefont {Tang}\ \emph {et~al.}(2018)\citenamefont {Tang},
  \citenamefont {Kao}, \citenamefont {Li}, \citenamefont {Seo}, \citenamefont
  {Mallayya}, \citenamefont {Rigol}, \citenamefont {Gopalakrishnan},\ and\
  \citenamefont {Lev}}]{tang2018}%
  \BibitemOpen
  \bibfield  {author} {\bibinfo {author} {\bibfnamefont {Y.}~\bibnamefont
  {Tang}}, \bibinfo {author} {\bibfnamefont {W.}~\bibnamefont {Kao}}, \bibinfo
  {author} {\bibfnamefont {K.-Y.}\ \bibnamefont {Li}}, \bibinfo {author}
  {\bibfnamefont {S.}~\bibnamefont {Seo}}, \bibinfo {author} {\bibfnamefont
  {K.}~\bibnamefont {Mallayya}}, \bibinfo {author} {\bibfnamefont
  {M.}~\bibnamefont {Rigol}}, \bibinfo {author} {\bibfnamefont
  {S.}~\bibnamefont {Gopalakrishnan}}, \ and\ \bibinfo {author} {\bibfnamefont
  {B.~L.}\ \bibnamefont {Lev}},\ }\href {\doibase 10.1103/PhysRevX.8.021030}
  {\bibfield  {journal} {\bibinfo  {journal} {Phys. Rev. X}\ }\textbf {\bibinfo
  {volume} {8}},\ \bibinfo {pages} {021030} (\bibinfo {year}
  {2018})}\BibitemShut {NoStop}%
\bibitem [{\citenamefont {Erne}\ \emph {et~al.}(2018)\citenamefont {Erne},
  \citenamefont {B{\"u}cker}, \citenamefont {Gasenzer}, \citenamefont
  {Berges},\ and\ \citenamefont {Schmiedmayer}}]{erne2018}%
  \BibitemOpen
  \bibfield  {author} {\bibinfo {author} {\bibfnamefont {S.}~\bibnamefont
  {Erne}}, \bibinfo {author} {\bibfnamefont {R.}~\bibnamefont {B{\"u}cker}},
  \bibinfo {author} {\bibfnamefont {T.}~\bibnamefont {Gasenzer}}, \bibinfo
  {author} {\bibfnamefont {J.}~\bibnamefont {Berges}}, \ and\ \bibinfo {author}
  {\bibfnamefont {J.}~\bibnamefont {Schmiedmayer}},\ }\href@noop {} {\bibfield
  {journal} {\bibinfo  {journal} {Nature}\ }\textbf {\bibinfo {volume} {563}},\
  \bibinfo {pages} {225} (\bibinfo {year} {2018})}\BibitemShut {NoStop}%
\bibitem [{\citenamefont {Zundel}\ \emph {et~al.}(2019)\citenamefont {Zundel},
  \citenamefont {Wilson}, \citenamefont {Malvania}, \citenamefont {Xia},
  \citenamefont {Riou},\ and\ \citenamefont {Weiss}}]{zundel2018}%
  \BibitemOpen
  \bibfield  {author} {\bibinfo {author} {\bibfnamefont {L.~A.}\ \bibnamefont
  {Zundel}}, \bibinfo {author} {\bibfnamefont {J.~M.}\ \bibnamefont {Wilson}},
  \bibinfo {author} {\bibfnamefont {N.}~\bibnamefont {Malvania}}, \bibinfo
  {author} {\bibfnamefont {L.}~\bibnamefont {Xia}}, \bibinfo {author}
  {\bibfnamefont {J.-F.}\ \bibnamefont {Riou}}, \ and\ \bibinfo {author}
  {\bibfnamefont {D.~S.}\ \bibnamefont {Weiss}},\ }\href {\doibase
  10.1103/PhysRevLett.122.013402} {\bibfield  {journal} {\bibinfo  {journal}
  {Phys. Rev. Lett.}\ }\textbf {\bibinfo {volume} {122}},\ \bibinfo {pages}
  {013402} (\bibinfo {year} {2019})}\BibitemShut {NoStop}%
\bibitem [{\citenamefont {Castro-Alvaredo}\ \emph {et~al.}(2016)\citenamefont
  {Castro-Alvaredo}, \citenamefont {Doyon},\ and\ \citenamefont
  {Yoshimura}}]{Doyon}%
  \BibitemOpen
  \bibfield  {author} {\bibinfo {author} {\bibfnamefont {O.~A.}\ \bibnamefont
  {Castro-Alvaredo}}, \bibinfo {author} {\bibfnamefont {B.}~\bibnamefont
  {Doyon}}, \ and\ \bibinfo {author} {\bibfnamefont {T.}~\bibnamefont
  {Yoshimura}},\ }\href {\doibase 10.1103/PhysRevX.6.041065} {\bibfield
  {journal} {\bibinfo  {journal} {Phys. Rev. X}\ }\textbf {\bibinfo {volume}
  {6}},\ \bibinfo {pages} {041065} (\bibinfo {year} {2016})}\BibitemShut
  {NoStop}%
\bibitem [{\citenamefont {Bertini}\ \emph {et~al.}(2016)\citenamefont
  {Bertini}, \citenamefont {Collura}, \citenamefont {De~Nardis},\ and\
  \citenamefont {Fagotti}}]{Fagotti}%
  \BibitemOpen
  \bibfield  {author} {\bibinfo {author} {\bibfnamefont {B.}~\bibnamefont
  {Bertini}}, \bibinfo {author} {\bibfnamefont {M.}~\bibnamefont {Collura}},
  \bibinfo {author} {\bibfnamefont {J.}~\bibnamefont {De~Nardis}}, \ and\
  \bibinfo {author} {\bibfnamefont {M.}~\bibnamefont {Fagotti}},\ }\href
  {\doibase 10.1103/PhysRevLett.117.207201} {\bibfield  {journal} {\bibinfo
  {journal} {Phys. Rev. Lett.}\ }\textbf {\bibinfo {volume} {117}},\ \bibinfo
  {pages} {207201} (\bibinfo {year} {2016})}\BibitemShut {NoStop}%
\bibitem [{\citenamefont {Doyon}\ and\ \citenamefont
  {Yoshimura}(2017)}]{SciPostPhys.2.2.014}%
  \BibitemOpen
  \bibfield  {author} {\bibinfo {author} {\bibfnamefont {B.}~\bibnamefont
  {Doyon}}\ and\ \bibinfo {author} {\bibfnamefont {T.}~\bibnamefont
  {Yoshimura}},\ }\href {\doibase 10.21468/SciPostPhys.2.2.014} {\bibfield
  {journal} {\bibinfo  {journal} {SciPost Phys.}\ }\textbf {\bibinfo {volume}
  {2}},\ \bibinfo {pages} {014} (\bibinfo {year} {2017})}\BibitemShut {NoStop}%
\bibitem [{\citenamefont {Ilievski}\ and\ \citenamefont
  {De~Nardis}(2017{\natexlab{a}})}]{PhysRevLett.119.020602}%
  \BibitemOpen
  \bibfield  {author} {\bibinfo {author} {\bibfnamefont {E.}~\bibnamefont
  {Ilievski}}\ and\ \bibinfo {author} {\bibfnamefont {J.}~\bibnamefont
  {De~Nardis}},\ }\href {\doibase 10.1103/PhysRevLett.119.020602} {\bibfield
  {journal} {\bibinfo  {journal} {Phys. Rev. Lett.}\ }\textbf {\bibinfo
  {volume} {119}},\ \bibinfo {pages} {020602} (\bibinfo {year}
  {2017}{\natexlab{a}})}\BibitemShut {NoStop}%
\bibitem [{\citenamefont {Bulchandani}\ \emph {et~al.}(2017)\citenamefont
  {Bulchandani}, \citenamefont {Vasseur}, \citenamefont {Karrasch},\ and\
  \citenamefont {Moore}}]{BBH0}%
  \BibitemOpen
  \bibfield  {author} {\bibinfo {author} {\bibfnamefont {V.~B.}\ \bibnamefont
  {Bulchandani}}, \bibinfo {author} {\bibfnamefont {R.}~\bibnamefont
  {Vasseur}}, \bibinfo {author} {\bibfnamefont {C.}~\bibnamefont {Karrasch}}, \
  and\ \bibinfo {author} {\bibfnamefont {J.~E.}\ \bibnamefont {Moore}},\ }\href
  {\doibase 10.1103/PhysRevLett.119.220604} {\bibfield  {journal} {\bibinfo
  {journal} {Phys. Rev. Lett.}\ }\textbf {\bibinfo {volume} {119}},\ \bibinfo
  {pages} {220604} (\bibinfo {year} {2017})}\BibitemShut {NoStop}%
\bibitem [{\citenamefont {Bulchandani}\ \emph {et~al.}(2018)\citenamefont
  {Bulchandani}, \citenamefont {Vasseur}, \citenamefont {Karrasch},\ and\
  \citenamefont {Moore}}]{BBH}%
  \BibitemOpen
  \bibfield  {author} {\bibinfo {author} {\bibfnamefont {V.~B.}\ \bibnamefont
  {Bulchandani}}, \bibinfo {author} {\bibfnamefont {R.}~\bibnamefont
  {Vasseur}}, \bibinfo {author} {\bibfnamefont {C.}~\bibnamefont {Karrasch}}, \
  and\ \bibinfo {author} {\bibfnamefont {J.~E.}\ \bibnamefont {Moore}},\ }\href
  {\doibase 10.1103/PhysRevB.97.045407} {\bibfield  {journal} {\bibinfo
  {journal} {Phys. Rev. B}\ }\textbf {\bibinfo {volume} {97}},\ \bibinfo
  {pages} {045407} (\bibinfo {year} {2018})}\BibitemShut {NoStop}%
\bibitem [{\citenamefont {Doyon}\ and\ \citenamefont
  {Spohn}(2017{\natexlab{a}})}]{GHDII}%
  \BibitemOpen
  \bibfield  {author} {\bibinfo {author} {\bibfnamefont {B.}~\bibnamefont
  {Doyon}}\ and\ \bibinfo {author} {\bibfnamefont {H.}~\bibnamefont {Spohn}},\
  }\href {\doibase 10.21468/SciPostPhys.3.6.039} {\bibfield  {journal}
  {\bibinfo  {journal} {SciPost Phys.}\ }\textbf {\bibinfo {volume} {3}},\
  \bibinfo {pages} {039} (\bibinfo {year} {2017}{\natexlab{a}})}\BibitemShut
  {NoStop}%
\bibitem [{\citenamefont {Doyon}\ and\ \citenamefont
  {Spohn}(2017{\natexlab{b}})}]{doyon2017dynamics}%
  \BibitemOpen
  \bibfield  {author} {\bibinfo {author} {\bibfnamefont {B.}~\bibnamefont
  {Doyon}}\ and\ \bibinfo {author} {\bibfnamefont {H.}~\bibnamefont {Spohn}},\
  }\href@noop {} {\bibfield  {journal} {\bibinfo  {journal} {Journal of
  Statistical Mechanics: Theory and Experiment}\ }\textbf {\bibinfo {volume}
  {2017}},\ \bibinfo {pages} {073210} (\bibinfo {year}
  {2017}{\natexlab{b}})}\BibitemShut {NoStop}%
\bibitem [{\citenamefont {Doyon}\ \emph {et~al.}(2018)\citenamefont {Doyon},
  \citenamefont {Yoshimura},\ and\ \citenamefont {Caux}}]{solitongases}%
  \BibitemOpen
  \bibfield  {author} {\bibinfo {author} {\bibfnamefont {B.}~\bibnamefont
  {Doyon}}, \bibinfo {author} {\bibfnamefont {T.}~\bibnamefont {Yoshimura}}, \
  and\ \bibinfo {author} {\bibfnamefont {J.-S.}\ \bibnamefont {Caux}},\ }\href
  {\doibase 10.1103/PhysRevLett.120.045301} {\bibfield  {journal} {\bibinfo
  {journal} {Phys. Rev. Lett.}\ }\textbf {\bibinfo {volume} {120}},\ \bibinfo
  {pages} {045301} (\bibinfo {year} {2018})}\BibitemShut {NoStop}%
\bibitem [{\citenamefont {Doyon}\ \emph {et~al.}(2017)\citenamefont {Doyon},
  \citenamefont {Dubail}, \citenamefont {Konik},\ and\ \citenamefont
  {Yoshimura}}]{PhysRevLett.119.195301}%
  \BibitemOpen
  \bibfield  {author} {\bibinfo {author} {\bibfnamefont {B.}~\bibnamefont
  {Doyon}}, \bibinfo {author} {\bibfnamefont {J.}~\bibnamefont {Dubail}},
  \bibinfo {author} {\bibfnamefont {R.}~\bibnamefont {Konik}}, \ and\ \bibinfo
  {author} {\bibfnamefont {T.}~\bibnamefont {Yoshimura}},\ }\href {\doibase
  10.1103/PhysRevLett.119.195301} {\bibfield  {journal} {\bibinfo  {journal}
  {Phys. Rev. Lett.}\ }\textbf {\bibinfo {volume} {119}},\ \bibinfo {pages}
  {195301} (\bibinfo {year} {2017})}\BibitemShut {NoStop}%
\bibitem [{\citenamefont {{Zotos}}(2016)}]{2016arXiv160408434Z}%
  \BibitemOpen
  \bibfield  {author} {\bibinfo {author} {\bibfnamefont {X.}~\bibnamefont
  {{Zotos}}},\ }\href@noop {} {\bibfield  {journal} {\bibinfo  {journal} {ArXiv
  e-prints}\ } (\bibinfo {year} {2016})},\ \Eprint
  {http://arxiv.org/abs/1604.08434} {arXiv:1604.08434 [cond-mat.stat-mech]}
  \BibitemShut {NoStop}%
\bibitem [{\citenamefont {Ilievski}\ and\ \citenamefont
  {De~Nardis}(2017{\natexlab{b}})}]{PhysRevB.96.081118}%
  \BibitemOpen
  \bibfield  {author} {\bibinfo {author} {\bibfnamefont {E.}~\bibnamefont
  {Ilievski}}\ and\ \bibinfo {author} {\bibfnamefont {J.}~\bibnamefont
  {De~Nardis}},\ }\href {\doibase 10.1103/PhysRevB.96.081118} {\bibfield
  {journal} {\bibinfo  {journal} {Phys. Rev. B}\ }\textbf {\bibinfo {volume}
  {96}},\ \bibinfo {pages} {081118} (\bibinfo {year}
  {2017}{\natexlab{b}})}\BibitemShut {NoStop}%
\bibitem [{\citenamefont {Collura}\ \emph {et~al.}(2018)\citenamefont
  {Collura}, \citenamefont {De~Luca},\ and\ \citenamefont
  {Viti}}]{PhysRevB.97.081111}%
  \BibitemOpen
  \bibfield  {author} {\bibinfo {author} {\bibfnamefont {M.}~\bibnamefont
  {Collura}}, \bibinfo {author} {\bibfnamefont {A.}~\bibnamefont {De~Luca}}, \
  and\ \bibinfo {author} {\bibfnamefont {J.}~\bibnamefont {Viti}},\ }\href
  {\doibase 10.1103/PhysRevB.97.081111} {\bibfield  {journal} {\bibinfo
  {journal} {Phys. Rev. B}\ }\textbf {\bibinfo {volume} {97}},\ \bibinfo
  {pages} {081111} (\bibinfo {year} {2018})}\BibitemShut {NoStop}%
\bibitem [{\citenamefont {De~Nardis}\ \emph {et~al.}(2018)\citenamefont
  {De~Nardis}, \citenamefont {Bernard},\ and\ \citenamefont {Doyon}}]{dbd1}%
  \BibitemOpen
  \bibfield  {author} {\bibinfo {author} {\bibfnamefont {J.}~\bibnamefont
  {De~Nardis}}, \bibinfo {author} {\bibfnamefont {D.}~\bibnamefont {Bernard}},
  \ and\ \bibinfo {author} {\bibfnamefont {B.}~\bibnamefont {Doyon}},\ }\href
  {\doibase 10.1103/PhysRevLett.121.160603} {\bibfield  {journal} {\bibinfo
  {journal} {Phys. Rev. Lett.}\ }\textbf {\bibinfo {volume} {121}},\ \bibinfo
  {pages} {160603} (\bibinfo {year} {2018})}\BibitemShut {NoStop}%
\bibitem [{\citenamefont {Gopalakrishnan}\ \emph {et~al.}(2018)\citenamefont
  {Gopalakrishnan}, \citenamefont {Huse}, \citenamefont {Khemani},\ and\
  \citenamefont {Vasseur}}]{ghkv}%
  \BibitemOpen
  \bibfield  {author} {\bibinfo {author} {\bibfnamefont {S.}~\bibnamefont
  {Gopalakrishnan}}, \bibinfo {author} {\bibfnamefont {D.~A.}\ \bibnamefont
  {Huse}}, \bibinfo {author} {\bibfnamefont {V.}~\bibnamefont {Khemani}}, \
  and\ \bibinfo {author} {\bibfnamefont {R.}~\bibnamefont {Vasseur}},\ }\href
  {\doibase 10.1103/PhysRevB.98.220303} {\bibfield  {journal} {\bibinfo
  {journal} {Phys. Rev. B}\ }\textbf {\bibinfo {volume} {98}},\ \bibinfo
  {pages} {220303} (\bibinfo {year} {2018})}\BibitemShut {NoStop}%
\bibitem [{\citenamefont {Nardis}\ \emph {et~al.}(2019)\citenamefont {Nardis},
  \citenamefont {Bernard},\ and\ \citenamefont {Doyon}}]{dbd2}%
  \BibitemOpen
  \bibfield  {author} {\bibinfo {author} {\bibfnamefont {J.~D.}\ \bibnamefont
  {Nardis}}, \bibinfo {author} {\bibfnamefont {D.}~\bibnamefont {Bernard}}, \
  and\ \bibinfo {author} {\bibfnamefont {B.}~\bibnamefont {Doyon}},\ }\href
  {\doibase 10.21468/SciPostPhys.6.4.049} {\bibfield  {journal} {\bibinfo
  {journal} {SciPost Phys.}\ }\textbf {\bibinfo {volume} {6}},\ \bibinfo
  {pages} {49} (\bibinfo {year} {2019})}\BibitemShut {NoStop}%
\bibitem [{\citenamefont {Gopalakrishnan}\ and\ \citenamefont
  {Vasseur}(2019)}]{gv_superdiffusion}%
  \BibitemOpen
  \bibfield  {author} {\bibinfo {author} {\bibfnamefont {S.}~\bibnamefont
  {Gopalakrishnan}}\ and\ \bibinfo {author} {\bibfnamefont {R.}~\bibnamefont
  {Vasseur}},\ }\href {\doibase 10.1103/PhysRevLett.122.127202} {\bibfield
  {journal} {\bibinfo  {journal} {Phys. Rev. Lett.}\ }\textbf {\bibinfo
  {volume} {122}},\ \bibinfo {pages} {127202} (\bibinfo {year}
  {2019})}\BibitemShut {NoStop}%
\bibitem [{\citenamefont {Agrawal}\ \emph {et~al.}(2019)\citenamefont
  {Agrawal}, \citenamefont {Gopalakrishnan},\ and\ \citenamefont
  {Vasseur}}]{agrawal2019}%
  \BibitemOpen
  \bibfield  {author} {\bibinfo {author} {\bibfnamefont {U.}~\bibnamefont
  {Agrawal}}, \bibinfo {author} {\bibfnamefont {S.}~\bibnamefont
  {Gopalakrishnan}}, \ and\ \bibinfo {author} {\bibfnamefont {R.}~\bibnamefont
  {Vasseur}},\ }\href {\doibase 10.1103/PhysRevB.99.174203} {\bibfield
  {journal} {\bibinfo  {journal} {Phys. Rev. B}\ }\textbf {\bibinfo {volume}
  {99}},\ \bibinfo {pages} {174203} (\bibinfo {year} {2019})}\BibitemShut
  {NoStop}%
\bibitem [{\citenamefont {Gopalakrishnan}\ \emph {et~al.}(2019)\citenamefont
  {Gopalakrishnan}, \citenamefont {Vasseur},\ and\ \citenamefont {Ware}}]{gvw}%
  \BibitemOpen
  \bibfield  {author} {\bibinfo {author} {\bibfnamefont {S.}~\bibnamefont
  {Gopalakrishnan}}, \bibinfo {author} {\bibfnamefont {R.}~\bibnamefont
  {Vasseur}}, \ and\ \bibinfo {author} {\bibfnamefont {B.}~\bibnamefont
  {Ware}},\ }\href {\doibase 10.1073/pnas.1906914116} {\bibfield  {journal}
  {\bibinfo  {journal} {Proceedings of the National Academy of Sciences}\
  }\textbf {\bibinfo {volume} {116}},\ \bibinfo {pages} {16250} (\bibinfo
  {year} {2019})}\BibitemShut {NoStop}%
\bibitem [{\citenamefont {Horvath}(2019)}]{horvath2019euler}%
  \BibitemOpen
  \bibfield  {author} {\bibinfo {author} {\bibfnamefont {D.~X.}\ \bibnamefont
  {Horvath}},\ }\href@noop {} {\bibfield  {journal} {\bibinfo  {journal} {arXiv
  preprint arXiv:1905.08590}\ } (\bibinfo {year} {2019})}\BibitemShut {NoStop}%
\bibitem [{\citenamefont {Bertini}\ \emph {et~al.}(2019)\citenamefont
  {Bertini}, \citenamefont {Piroli},\ and\ \citenamefont
  {Kormos}}]{PhysRevB.100.035108}%
  \BibitemOpen
  \bibfield  {author} {\bibinfo {author} {\bibfnamefont {B.}~\bibnamefont
  {Bertini}}, \bibinfo {author} {\bibfnamefont {L.}~\bibnamefont {Piroli}}, \
  and\ \bibinfo {author} {\bibfnamefont {M.}~\bibnamefont {Kormos}},\ }\href
  {\doibase 10.1103/PhysRevB.100.035108} {\bibfield  {journal} {\bibinfo
  {journal} {Phys. Rev. B}\ }\textbf {\bibinfo {volume} {100}},\ \bibinfo
  {pages} {035108} (\bibinfo {year} {2019})}\BibitemShut {NoStop}%
\bibitem [{\citenamefont {{Bastianello}}\ \emph {et~al.}(2019)\citenamefont
  {{Bastianello}}, \citenamefont {{Alba}},\ and\ \citenamefont {{S{\'e}bastien
  Caux}}}]{2019arXiv190601654B}%
  \BibitemOpen
  \bibfield  {author} {\bibinfo {author} {\bibfnamefont {A.}~\bibnamefont
  {{Bastianello}}}, \bibinfo {author} {\bibfnamefont {V.}~\bibnamefont
  {{Alba}}}, \ and\ \bibinfo {author} {\bibfnamefont {J.}~\bibnamefont
  {{S{\'e}bastien Caux}}},\ }\href@noop {} {\bibfield  {journal} {\bibinfo
  {journal} {arXiv e-prints}\ ,\ \bibinfo {eid} {arXiv:1906.01654}} (\bibinfo
  {year} {2019})},\ \Eprint {http://arxiv.org/abs/1906.01654} {arXiv:1906.01654
  [cond-mat.stat-mech]} \BibitemShut {NoStop}%
\bibitem [{\citenamefont {Zotos}(1999)}]{PhysRevLett.82.1764}%
  \BibitemOpen
  \bibfield  {author} {\bibinfo {author} {\bibfnamefont {X.}~\bibnamefont
  {Zotos}},\ }\href {\doibase 10.1103/PhysRevLett.82.1764} {\bibfield
  {journal} {\bibinfo  {journal} {Phys. Rev. Lett.}\ }\textbf {\bibinfo
  {volume} {82}},\ \bibinfo {pages} {1764} (\bibinfo {year}
  {1999})}\BibitemShut {NoStop}%
\bibitem [{\citenamefont {Prosen}\ and\ \citenamefont
  {Ilievski}(2013)}]{PhysRevLett.111.057203}%
  \BibitemOpen
  \bibfield  {author} {\bibinfo {author} {\bibfnamefont {T.}~\bibnamefont
  {Prosen}}\ and\ \bibinfo {author} {\bibfnamefont {E.}~\bibnamefont
  {Ilievski}},\ }\href {\doibase 10.1103/PhysRevLett.111.057203} {\bibfield
  {journal} {\bibinfo  {journal} {Phys. Rev. Lett.}\ }\textbf {\bibinfo
  {volume} {111}},\ \bibinfo {pages} {057203} (\bibinfo {year}
  {2013})}\BibitemShut {NoStop}%
\bibitem [{\citenamefont {Medenjak}\ \emph {et~al.}(2017)\citenamefont
  {Medenjak}, \citenamefont {Karrasch},\ and\ \citenamefont {Prosen}}]{mkp}%
  \BibitemOpen
  \bibfield  {author} {\bibinfo {author} {\bibfnamefont {M.}~\bibnamefont
  {Medenjak}}, \bibinfo {author} {\bibfnamefont {C.}~\bibnamefont {Karrasch}},
  \ and\ \bibinfo {author} {\bibfnamefont {T.}~\bibnamefont {Prosen}},\ }\href
  {\doibase 10.1103/PhysRevLett.119.080602} {\bibfield  {journal} {\bibinfo
  {journal} {Phys. Rev. Lett.}\ }\textbf {\bibinfo {volume} {119}},\ \bibinfo
  {pages} {080602} (\bibinfo {year} {2017})}\BibitemShut {NoStop}%
\bibitem [{\citenamefont {Carmelo}\ and\ \citenamefont
  {Prosen}(2017)}]{carmelo2017}%
  \BibitemOpen
  \bibfield  {author} {\bibinfo {author} {\bibfnamefont {J.~M.~P.}\
  \bibnamefont {Carmelo}}\ and\ \bibinfo {author} {\bibfnamefont
  {T.}~\bibnamefont {Prosen}},\ }\href@noop {} {\bibfield  {journal} {\bibinfo
  {journal} {Nuclear Physics B}\ }\textbf {\bibinfo {volume} {914}},\ \bibinfo
  {pages} {62} (\bibinfo {year} {2017})}\BibitemShut {NoStop}%
\bibitem [{\citenamefont {Sirker}\ \emph {et~al.}(2009)\citenamefont {Sirker},
  \citenamefont {Pereira},\ and\ \citenamefont {Affleck}}]{sirker:2010}%
  \BibitemOpen
  \bibfield  {author} {\bibinfo {author} {\bibfnamefont {J.}~\bibnamefont
  {Sirker}}, \bibinfo {author} {\bibfnamefont {R.~G.}\ \bibnamefont {Pereira}},
  \ and\ \bibinfo {author} {\bibfnamefont {I.}~\bibnamefont {Affleck}},\ }\href
  {\doibase 10.1103/PhysRevLett.103.216602} {\bibfield  {journal} {\bibinfo
  {journal} {Phys. Rev. Lett.}\ }\textbf {\bibinfo {volume} {103}},\ \bibinfo
  {pages} {216602} (\bibinfo {year} {2009})}\BibitemShut {NoStop}%
\bibitem [{\citenamefont {Pereira}\ \emph {et~al.}(2014)\citenamefont
  {Pereira}, \citenamefont {Pasquier}, \citenamefont {Sirker},\ and\
  \citenamefont {Affleck}}]{1742-5468-2014-9-P09037}%
  \BibitemOpen
  \bibfield  {author} {\bibinfo {author} {\bibfnamefont {R.~G.}\ \bibnamefont
  {Pereira}}, \bibinfo {author} {\bibfnamefont {V.}~\bibnamefont {Pasquier}},
  \bibinfo {author} {\bibfnamefont {J.}~\bibnamefont {Sirker}}, \ and\ \bibinfo
  {author} {\bibfnamefont {I.}~\bibnamefont {Affleck}},\ }\href
  {http://stacks.iop.org/1742-5468/2014/i=9/a=P09037} {\bibfield  {journal}
  {\bibinfo  {journal} {Journal of Statistical Mechanics: Theory and
  Experiment}\ }\textbf {\bibinfo {volume} {2014}},\ \bibinfo {pages} {P09037}
  (\bibinfo {year} {2014})}\BibitemShut {NoStop}%
\bibitem [{\citenamefont {Urichuk}\ \emph {et~al.}(2019)\citenamefont
  {Urichuk}, \citenamefont {Oez}, \citenamefont {Kl{\"u}mper},\ and\
  \citenamefont {Sirker}}]{urichuk2019spin}%
  \BibitemOpen
  \bibfield  {author} {\bibinfo {author} {\bibfnamefont {A.}~\bibnamefont
  {Urichuk}}, \bibinfo {author} {\bibfnamefont {Y.}~\bibnamefont {Oez}},
  \bibinfo {author} {\bibfnamefont {A.}~\bibnamefont {Kl{\"u}mper}}, \ and\
  \bibinfo {author} {\bibfnamefont {J.}~\bibnamefont {Sirker}},\ }\href@noop {}
  {\bibfield  {journal} {\bibinfo  {journal} {SciPost Physics}\ }\textbf
  {\bibinfo {volume} {6}},\ \bibinfo {pages} {005} (\bibinfo {year}
  {2019})}\BibitemShut {NoStop}%
\bibitem [{\citenamefont {Ljubotina}\ \emph {et~al.}(2017)\citenamefont
  {Ljubotina}, \citenamefont {{\v Z}nidari{\v c}},\ and\ \citenamefont
  {Prosen}}]{lzp}%
  \BibitemOpen
  \bibfield  {author} {\bibinfo {author} {\bibfnamefont {M.}~\bibnamefont
  {Ljubotina}}, \bibinfo {author} {\bibfnamefont {M.}~\bibnamefont {{\v
  Z}nidari{\v c}}}, \ and\ \bibinfo {author} {\bibfnamefont {T.}~\bibnamefont
  {Prosen}},\ }\href {http://dx.doi.org/10.1038/ncomms16117} {\bibfield
  {journal} {\bibinfo  {journal} {Nature Communications}\ }\textbf {\bibinfo
  {volume} {8}},\ \bibinfo {pages} {16117 EP } (\bibinfo {year}
  {2017})}\BibitemShut {NoStop}%
\bibitem [{\citenamefont {Ljubotina}\ \emph {et~al.}(2019)\citenamefont
  {Ljubotina}, \citenamefont {\ifmmode \check{Z}\else
  \v{Z}\fi{}nidari\ifmmode~\check{c}\else \v{c}\fi{}},\ and\ \citenamefont
  {Prosen}}]{PhysRevLett.122.210602}%
  \BibitemOpen
  \bibfield  {author} {\bibinfo {author} {\bibfnamefont {M.}~\bibnamefont
  {Ljubotina}}, \bibinfo {author} {\bibfnamefont {M.}~\bibnamefont {\ifmmode
  \check{Z}\else \v{Z}\fi{}nidari\ifmmode~\check{c}\else \v{c}\fi{}}}, \ and\
  \bibinfo {author} {\bibfnamefont {T.}~\bibnamefont {Prosen}},\ }\href
  {\doibase 10.1103/PhysRevLett.122.210602} {\bibfield  {journal} {\bibinfo
  {journal} {Phys. Rev. Lett.}\ }\textbf {\bibinfo {volume} {122}},\ \bibinfo
  {pages} {210602} (\bibinfo {year} {2019})}\BibitemShut {NoStop}%
\bibitem [{\citenamefont {Dupont}\ and\ \citenamefont
  {Moore}(2019)}]{dupont_moore}%
  \BibitemOpen
  \bibfield  {author} {\bibinfo {author} {\bibfnamefont {M.}~\bibnamefont
  {Dupont}}\ and\ \bibinfo {author} {\bibfnamefont {J.~E.}\ \bibnamefont
  {Moore}},\ }\href@noop {} {\bibfield  {journal} {\bibinfo  {journal} {arXiv
  preprint arXiv:1907.12115}\ } (\bibinfo {year} {2019})}\BibitemShut {NoStop}%
\bibitem [{\citenamefont {Weiner}\ \emph {et~al.}(2019)\citenamefont {Weiner},
  \citenamefont {Schmitteckert}, \citenamefont {Bera},\ and\ \citenamefont
  {Evers}}]{weiner2019}%
  \BibitemOpen
  \bibfield  {author} {\bibinfo {author} {\bibfnamefont {F.}~\bibnamefont
  {Weiner}}, \bibinfo {author} {\bibfnamefont {P.}~\bibnamefont
  {Schmitteckert}}, \bibinfo {author} {\bibfnamefont {S.}~\bibnamefont {Bera}},
  \ and\ \bibinfo {author} {\bibfnamefont {F.}~\bibnamefont {Evers}},\
  }\href@noop {} {\bibfield  {journal} {\bibinfo  {journal} {arXiv:1908.11432}\
  } (\bibinfo {year} {2019})}\BibitemShut {NoStop}%
\bibitem [{\citenamefont {Sachdev}\ and\ \citenamefont
  {Damle}(1997)}]{PhysRevLett.78.943}%
  \BibitemOpen
  \bibfield  {author} {\bibinfo {author} {\bibfnamefont {S.}~\bibnamefont
  {Sachdev}}\ and\ \bibinfo {author} {\bibfnamefont {K.}~\bibnamefont
  {Damle}},\ }\href {\doibase 10.1103/PhysRevLett.78.943} {\bibfield  {journal}
  {\bibinfo  {journal} {Phys. Rev. Lett.}\ }\textbf {\bibinfo {volume} {78}},\
  \bibinfo {pages} {943} (\bibinfo {year} {1997})}\BibitemShut {NoStop}%
\bibitem [{\citenamefont {Damle}\ and\ \citenamefont
  {Sachdev}(1998)}]{PhysRevB.57.8307}%
  \BibitemOpen
  \bibfield  {author} {\bibinfo {author} {\bibfnamefont {K.}~\bibnamefont
  {Damle}}\ and\ \bibinfo {author} {\bibfnamefont {S.}~\bibnamefont
  {Sachdev}},\ }\href {\doibase 10.1103/PhysRevB.57.8307} {\bibfield  {journal}
  {\bibinfo  {journal} {Phys. Rev. B}\ }\textbf {\bibinfo {volume} {57}},\
  \bibinfo {pages} {8307} (\bibinfo {year} {1998})}\BibitemShut {NoStop}%
\bibitem [{\citenamefont {Damle}\ and\ \citenamefont
  {Sachdev}(2005)}]{PhysRevLett.95.187201}%
  \BibitemOpen
  \bibfield  {author} {\bibinfo {author} {\bibfnamefont {K.}~\bibnamefont
  {Damle}}\ and\ \bibinfo {author} {\bibfnamefont {S.}~\bibnamefont
  {Sachdev}},\ }\href {\doibase 10.1103/PhysRevLett.95.187201} {\bibfield
  {journal} {\bibinfo  {journal} {Phys. Rev. Lett.}\ }\textbf {\bibinfo
  {volume} {95}},\ \bibinfo {pages} {187201} (\bibinfo {year}
  {2005})}\BibitemShut {NoStop}%
\bibitem [{\citenamefont {Ilievski}\ \emph {et~al.}(2018)\citenamefont
  {Ilievski}, \citenamefont {De~Nardis}, \citenamefont {Medenjak},\ and\
  \citenamefont {Prosen}}]{idmp}%
  \BibitemOpen
  \bibfield  {author} {\bibinfo {author} {\bibfnamefont {E.}~\bibnamefont
  {Ilievski}}, \bibinfo {author} {\bibfnamefont {J.}~\bibnamefont {De~Nardis}},
  \bibinfo {author} {\bibfnamefont {M.}~\bibnamefont {Medenjak}}, \ and\
  \bibinfo {author} {\bibfnamefont {T.}~\bibnamefont {Prosen}},\ }\href
  {\doibase 10.1103/PhysRevLett.121.230602} {\bibfield  {journal} {\bibinfo
  {journal} {Phys. Rev. Lett.}\ }\textbf {\bibinfo {volume} {121}},\ \bibinfo
  {pages} {230602} (\bibinfo {year} {2018})}\BibitemShut {NoStop}%
\bibitem [{\citenamefont {De~Nardis}\ \emph {et~al.}(2019)\citenamefont
  {De~Nardis}, \citenamefont {Medenjak}, \citenamefont {Karrasch},\ and\
  \citenamefont {Ilievski}}]{denardis_superdiffusion}%
  \BibitemOpen
  \bibfield  {author} {\bibinfo {author} {\bibfnamefont {J.}~\bibnamefont
  {De~Nardis}}, \bibinfo {author} {\bibfnamefont {M.}~\bibnamefont {Medenjak}},
  \bibinfo {author} {\bibfnamefont {C.}~\bibnamefont {Karrasch}}, \ and\
  \bibinfo {author} {\bibfnamefont {E.}~\bibnamefont {Ilievski}},\ }\href@noop
  {} {\bibfield  {journal} {\bibinfo  {journal} {arXiv preprint
  arXiv:1903.07598}\ } (\bibinfo {year} {2019})}\BibitemShut {NoStop}%
\bibitem [{\citenamefont {Prosen}(2014)}]{Prosen20141177}%
  \BibitemOpen
  \bibfield  {author} {\bibinfo {author} {\bibfnamefont {T.}~\bibnamefont
  {Prosen}},\ }\href {\doibase
  http://dx.doi.org/10.1016/j.nuclphysb.2014.07.024} {\bibfield  {journal}
  {\bibinfo  {journal} {Nuclear Physics B}\ }\textbf {\bibinfo {volume}
  {886}},\ \bibinfo {pages} {1177 } (\bibinfo {year} {2014})}\BibitemShut
  {NoStop}%
\bibitem [{\citenamefont {S{\'a}nchez}\ \emph {et~al.}(2018)\citenamefont
  {S{\'a}nchez}, \citenamefont {Varma},\ and\ \citenamefont
  {Oganesyan}}]{sanchez2018anomalous}%
  \BibitemOpen
  \bibfield  {author} {\bibinfo {author} {\bibfnamefont {R.~J.}\ \bibnamefont
  {S{\'a}nchez}}, \bibinfo {author} {\bibfnamefont {V.~K.}\ \bibnamefont
  {Varma}}, \ and\ \bibinfo {author} {\bibfnamefont {V.}~\bibnamefont
  {Oganesyan}},\ }\href@noop {} {\bibfield  {journal} {\bibinfo  {journal}
  {Physical Review B}\ }\textbf {\bibinfo {volume} {98}},\ \bibinfo {pages}
  {054415} (\bibinfo {year} {2018})}\BibitemShut {NoStop}%
\bibitem [{\citenamefont {Lieb}\ and\ \citenamefont
  {Robinson}(2004)}]{Lieb2004}%
  \BibitemOpen
  \bibfield  {author} {\bibinfo {author} {\bibfnamefont {E.~H.}\ \bibnamefont
  {Lieb}}\ and\ \bibinfo {author} {\bibfnamefont {D.~W.}\ \bibnamefont
  {Robinson}},\ }\enquote {\bibinfo {title} {The finite group velocity of
  quantum spin systems},}\ in\ \href {\doibase 10.1007/978-3-662-10018-9_25}
  {\emph {\bibinfo {booktitle} {Statistical Mechanics: Selecta of Elliott H.
  Lieb}}},\ \bibinfo {editor} {edited by\ \bibinfo {editor} {\bibfnamefont
  {B.}~\bibnamefont {Nachtergaele}}, \bibinfo {editor} {\bibfnamefont {J.~P.}\
  \bibnamefont {Solovej}}, \ and\ \bibinfo {editor} {\bibfnamefont
  {J.}~\bibnamefont {Yngvason}}}\ (\bibinfo  {publisher} {Springer Berlin
  Heidelberg},\ \bibinfo {address} {Berlin, Heidelberg},\ \bibinfo {year}
  {2004})\ pp.\ \bibinfo {pages} {425--431}\BibitemShut {NoStop}%
\bibitem [{\citenamefont {Rigol}\ \emph {et~al.}(2007)\citenamefont {Rigol},
  \citenamefont {Dunjko}, \citenamefont {Yurovsky},\ and\ \citenamefont
  {Olshanii}}]{PhysRevLett.98.050405}%
  \BibitemOpen
  \bibfield  {author} {\bibinfo {author} {\bibfnamefont {M.}~\bibnamefont
  {Rigol}}, \bibinfo {author} {\bibfnamefont {V.}~\bibnamefont {Dunjko}},
  \bibinfo {author} {\bibfnamefont {V.}~\bibnamefont {Yurovsky}}, \ and\
  \bibinfo {author} {\bibfnamefont {M.}~\bibnamefont {Olshanii}},\ }\href
  {\doibase 10.1103/PhysRevLett.98.050405} {\bibfield  {journal} {\bibinfo
  {journal} {Phys. Rev. Lett.}\ }\textbf {\bibinfo {volume} {98}},\ \bibinfo
  {pages} {050405} (\bibinfo {year} {2007})}\BibitemShut {NoStop}%
\bibitem [{\citenamefont {Caux}\ and\ \citenamefont
  {Konik}(2012)}]{caux_konik}%
  \BibitemOpen
  \bibfield  {author} {\bibinfo {author} {\bibfnamefont {J.-S.}\ \bibnamefont
  {Caux}}\ and\ \bibinfo {author} {\bibfnamefont {R.~M.}\ \bibnamefont
  {Konik}},\ }\href {\doibase 10.1103/PhysRevLett.109.175301} {\bibfield
  {journal} {\bibinfo  {journal} {Phys. Rev. Lett.}\ }\textbf {\bibinfo
  {volume} {109}},\ \bibinfo {pages} {175301} (\bibinfo {year}
  {2012})}\BibitemShut {NoStop}%
\bibitem [{\citenamefont {Fagotti}\ and\ \citenamefont
  {Essler}(2013)}]{1742-5468-2013-07-P07012}%
  \BibitemOpen
  \bibfield  {author} {\bibinfo {author} {\bibfnamefont {M.}~\bibnamefont
  {Fagotti}}\ and\ \bibinfo {author} {\bibfnamefont {F.~H.~L.}\ \bibnamefont
  {Essler}},\ }\href {http://stacks.iop.org/1742-5468/2013/i=07/a=P07012}
  {\bibfield  {journal} {\bibinfo  {journal} {Journal of Statistical Mechanics:
  Theory and Experiment}\ }\textbf {\bibinfo {volume} {2013}},\ \bibinfo
  {pages} {P07012} (\bibinfo {year} {2013})}\BibitemShut {NoStop}%
\bibitem [{\citenamefont {Pozsgay}(2013)}]{1742-5468-2013-07-P07003}%
  \BibitemOpen
  \bibfield  {author} {\bibinfo {author} {\bibfnamefont {B.}~\bibnamefont
  {Pozsgay}},\ }\href {http://stacks.iop.org/1742-5468/2013/i=07/a=P07003}
  {\bibfield  {journal} {\bibinfo  {journal} {Journal of Statistical Mechanics:
  Theory and Experiment}\ }\textbf {\bibinfo {volume} {2013}},\ \bibinfo
  {pages} {P07003} (\bibinfo {year} {2013})}\BibitemShut {NoStop}%
\bibitem [{\citenamefont {Pozsgay}\ \emph {et~al.}(2017)\citenamefont
  {Pozsgay}, \citenamefont {Vernier},\ and\ \citenamefont {Werner}}]{Pozsgay}%
  \BibitemOpen
  \bibfield  {author} {\bibinfo {author} {\bibfnamefont {B.}~\bibnamefont
  {Pozsgay}}, \bibinfo {author} {\bibfnamefont {E.}~\bibnamefont {Vernier}}, \
  and\ \bibinfo {author} {\bibfnamefont {M.~A.}\ \bibnamefont {Werner}},\
  }\href {http://arxiv.org/abs/1703.09516} {\  (\bibinfo {year} {2017})},\
  \Eprint {http://arxiv.org/abs/1703.09516} {arXiv:1703.09516} \BibitemShut
  {NoStop}%
\bibitem [{\citenamefont {El}(2003)}]{el2003thermodynamic}%
  \BibitemOpen
  \bibfield  {author} {\bibinfo {author} {\bibfnamefont {G.}~\bibnamefont
  {El}},\ }\href@noop {} {\bibfield  {journal} {\bibinfo  {journal} {Physics
  Letters A}\ }\textbf {\bibinfo {volume} {311}},\ \bibinfo {pages} {374}
  (\bibinfo {year} {2003})}\BibitemShut {NoStop}%
\bibitem [{\citenamefont {Gopalakrishnan}(2018)}]{sg_ffa}%
  \BibitemOpen
  \bibfield  {author} {\bibinfo {author} {\bibfnamefont {S.}~\bibnamefont
  {Gopalakrishnan}},\ }\href {\doibase 10.1103/PhysRevB.98.060302} {\bibfield
  {journal} {\bibinfo  {journal} {Phys. Rev. B}\ }\textbf {\bibinfo {volume}
  {98}},\ \bibinfo {pages} {060302} (\bibinfo {year} {2018})}\BibitemShut
  {NoStop}%
\bibitem [{sup()}]{suppmat}%
  \BibitemOpen
  \href@noop {} {\bibinfo  {journal} {See {S}upplemental {I}nformation for
  Thermodynamic Bethe Ansatz and GHD formulas, and details of numerical MPO
  calculations}\ }\BibitemShut {NoStop}%
\bibitem [{\citenamefont {Bernik}\ \emph {et~al.}(2013)\citenamefont {Bernik},
  \citenamefont {Beresnevich}, \citenamefont {Goetze},\ and\ \citenamefont
  {Kukso}}]{bernik2013distribution}%
  \BibitemOpen
\bibfield  {journal} {  }\bibfield  {author} {\bibinfo {author} {\bibfnamefont
  {V.}~\bibnamefont {Bernik}}, \bibinfo {author} {\bibfnamefont
  {V.}~\bibnamefont {Beresnevich}}, \bibinfo {author} {\bibfnamefont
  {F.}~\bibnamefont {Goetze}}, \ and\ \bibinfo {author} {\bibfnamefont
  {O.}~\bibnamefont {Kukso}},\ }in\ \href@noop {} {\emph {\bibinfo {booktitle}
  {Limit Theorems in Probability, Statistics and Number Theory}}}\ (\bibinfo
  {publisher} {Springer},\ \bibinfo {year} {2013})\ pp.\ \bibinfo {pages}
  {23--48}\BibitemShut {NoStop}%
\bibitem [{\citenamefont {{Van Damme}}\ \emph {et~al.}(2019)\citenamefont {{Van
  Damme}}, \citenamefont {{Vanderstraeten}}, \citenamefont {{De Nardis}},
  \citenamefont {{Haegeman}},\ and\ \citenamefont
  {{Verstraete}}}]{2019arXiv190702474V}%
  \BibitemOpen
  \bibfield  {author} {\bibinfo {author} {\bibfnamefont {M.}~\bibnamefont {{Van
  Damme}}}, \bibinfo {author} {\bibfnamefont {L.}~\bibnamefont
  {{Vanderstraeten}}}, \bibinfo {author} {\bibfnamefont {J.}~\bibnamefont {{De
  Nardis}}}, \bibinfo {author} {\bibfnamefont {J.}~\bibnamefont {{Haegeman}}},
  \ and\ \bibinfo {author} {\bibfnamefont {F.}~\bibnamefont {{Verstraete}}},\
  }\href@noop {} {\bibfield  {journal} {\bibinfo  {journal} {arXiv e-prints}\
  ,\ \bibinfo {eid} {arXiv:1907.02474}} (\bibinfo {year} {2019})},\ \Eprint
  {http://arxiv.org/abs/1907.02474} {arXiv:1907.02474 [cond-mat.str-el]}
  \BibitemShut {NoStop}%
\bibitem [{\citenamefont {Bouchaud}\ and\ \citenamefont
  {Georges}(1990)}]{bouchaud_georges}%
  \BibitemOpen
  \bibfield  {author} {\bibinfo {author} {\bibfnamefont {J.-P.}\ \bibnamefont
  {Bouchaud}}\ and\ \bibinfo {author} {\bibfnamefont {A.}~\bibnamefont
  {Georges}},\ }\href@noop {} {\bibfield  {journal} {\bibinfo  {journal}
  {Physics reports}\ }\textbf {\bibinfo {volume} {195}},\ \bibinfo {pages}
  {127} (\bibinfo {year} {1990})}\BibitemShut {NoStop}%
\bibitem [{\citenamefont {Vidal}(2003)}]{PhysRevLett.91.147902}%
  \BibitemOpen
  \bibfield  {author} {\bibinfo {author} {\bibfnamefont {G.}~\bibnamefont
  {Vidal}},\ }\href {\doibase 10.1103/PhysRevLett.91.147902} {\bibfield
  {journal} {\bibinfo  {journal} {Phys. Rev. Lett.}\ }\textbf {\bibinfo
  {volume} {91}},\ \bibinfo {pages} {147902} (\bibinfo {year}
  {2003})}\BibitemShut {NoStop}%
\bibitem [{\citenamefont {Zwolak}\ and\ \citenamefont {Vidal}(2004)}]{vidal}%
  \BibitemOpen
  \bibfield  {author} {\bibinfo {author} {\bibfnamefont {M.}~\bibnamefont
  {Zwolak}}\ and\ \bibinfo {author} {\bibfnamefont {G.}~\bibnamefont {Vidal}},\
  }\href@noop {} {\bibfield  {journal} {\bibinfo  {journal} {Phys. Rev. Lett.}\
  }\textbf {\bibinfo {volume} {93}},\ \bibinfo {pages} {207205} (\bibinfo
  {year} {2004})}\BibitemShut {NoStop}%
\bibitem [{\citenamefont {White}\ and\ \citenamefont
  {Feiguin}(2004)}]{whitetdmrg}%
  \BibitemOpen
  \bibfield  {author} {\bibinfo {author} {\bibfnamefont {S.~R.}\ \bibnamefont
  {White}}\ and\ \bibinfo {author} {\bibfnamefont {A.}~\bibnamefont
  {Feiguin}},\ }\href@noop {} {\bibfield  {journal} {\bibinfo  {journal} {Phys.
  Rev. Lett.}\ }\textbf {\bibinfo {volume} {93}},\ \bibinfo {pages} {076401}
  (\bibinfo {year} {2004})}\BibitemShut {NoStop}%
\bibitem [{\citenamefont {Verstraete}\ \emph {et~al.}(2004)\citenamefont
  {Verstraete}, \citenamefont {Garc\'{\i}a-Ripoll},\ and\ \citenamefont
  {Cirac}}]{vgrc}%
  \BibitemOpen
  \bibfield  {author} {\bibinfo {author} {\bibfnamefont {F.}~\bibnamefont
  {Verstraete}}, \bibinfo {author} {\bibfnamefont {J.~J.}\ \bibnamefont
  {Garc\'{\i}a-Ripoll}}, \ and\ \bibinfo {author} {\bibfnamefont {J.~I.}\
  \bibnamefont {Cirac}},\ }\href {\doibase 10.1103/PhysRevLett.93.207204}
  {\bibfield  {journal} {\bibinfo  {journal} {Phys. Rev. Lett.}\ }\textbf
  {\bibinfo {volume} {93}},\ \bibinfo {pages} {207204} (\bibinfo {year}
  {2004})}\BibitemShut {NoStop}%
\bibitem [{\citenamefont {Schollwoeck}(2011)}]{schollwoeck}%
  \BibitemOpen
  \bibfield  {author} {\bibinfo {author} {\bibfnamefont {U.}~\bibnamefont
  {Schollwoeck}},\ }\href {\doibase
  http://dx.doi.org/10.1016/j.aop.2010.09.012} {\bibfield  {journal} {\bibinfo
  {journal} {Annals of Physics}\ }\textbf {\bibinfo {volume} {326}},\ \bibinfo
  {pages} {96 } (\bibinfo {year} {2011})},\ \bibinfo {note} {january 2011
  Special Issue}\BibitemShut {NoStop}%
\bibitem [{\citenamefont {\ifmmode \check{Z}\else
  \v{Z}\fi{}nidari\ifmmode~\check{c}\else
  \v{c}\fi{}}(2011)}]{PhysRevLett.106.220601}%
  \BibitemOpen
  \bibfield  {author} {\bibinfo {author} {\bibfnamefont {M.}~\bibnamefont
  {\ifmmode \check{Z}\else \v{Z}\fi{}nidari\ifmmode~\check{c}\else
  \v{c}\fi{}}},\ }\href {\doibase 10.1103/PhysRevLett.106.220601} {\bibfield
  {journal} {\bibinfo  {journal} {Phys. Rev. Lett.}\ }\textbf {\bibinfo
  {volume} {106}},\ \bibinfo {pages} {220601} (\bibinfo {year}
  {2011})}\BibitemShut {NoStop}%
\end{thebibliography}%

%

%

\bigskip

\onecolumngrid
\newpage



\section*{Supplementary material}

# Supplemental Material for "Anomalous low-frequency conductivity in easy-plane XXZ spin chains"

Utkarsh Agrawal[1], Sarang Gopalakrishnan[2], Romain Vasseur[1], and Brayden Ware[1,3]

[1] *Department of Physics, University of Massachusetts, Amherst, MA 01003, USA*

[2] *Department of Physics and Astronomy, CUNY College of Staten Island,*
*Staten Island, NY 10314; Physics Program and Initiative for the Theoretical Sciences,*
*The Graduate Center, CUNY, New York, NY 10016, USA and*

[3] *Rudolf Peierls Center for Theoretical Physics, Clarendon Laboratory, Oxford OX1 3PU, UK*

## I. TEBD SIMULATIONS

We compute the current correlation function $C(x - x', t) = \langle j_x(t) j_{x'}(0) \rangle$ and the zero-momentum correlator

$$C(t) = \sum_x C(x, t) = \langle J(t) j_{x'}(0) \rangle, \quad J = \sum_x j_x$$

using a single time-evolution of the local operator $j_{x=0}$ in the Heisenberg picture as a matrix product operator (MPO). $C(t)$ is effectively computed in the infinite size limit by working with a large system size with boundaries well outside the lightcone of the operator for all times reached by the simulations. This allows us to take advantage of translation symmetry, using the single evolution of the local operator $j_{x=0}(t)$ rather than the more expensive evolution of the non-local operator $J(t)$ or many independent simulations $j_x(t)$ for each $x$. Since the MPO tensors of $j_{x=0}(t)$ only differ significantly from those of the identity operator inside a finite sized lightcone, we can construct an MPO for $j_x(t)$ without worrying about boundary effects by just shifting the non-trivial tensors of $j_0(t)$ and padding on either side with the tensors of the identity. In addition, we can take advantage of time-reversal symmetry $j(-t) = -j^*(t)$ and time-translation invariance by computing instead $C(x, t) = \langle j_x(t/2) j_0(-t/2) \rangle = - \langle j_x(t/2) j_0^*(t/2) \rangle$.

The TEBD simulation is run using a fourth-order Trotter decomposition with time step $\delta t = 0.1$. Truncations are done initially with a fixed truncation error $\varepsilon = 10^{-10}$, where the bond dimension grows rapidly near the center until a maximum of $\chi = 512$ - and then we continue with this fixed maximum bond dimension. We check for convergence of $C(t)$ by varying the maximum bond dimension as shown in Fig. 1a.

Previous works have used the technique of linear prediction, which effectively models the signal with a finite sum of decaying and oscillating exponentials, see *e.g.* [1, 2]. We find that we can convincingly extend the time series of $C(t)$ for the XXZ chain with $\Delta = \frac{1}{2}$ ($\lambda = \frac{1}{3}$), as shown in Fig. 1b, but the technique is unsuccessful for other values of $\Delta$, compatible with our expectations. We check the accuracy of the technique by leaving the infinite time limit of $C(t)$ as an undetermined parameter, and then comparing to the theoretical predicted Drude weight. For $\Delta = 0.5$, the

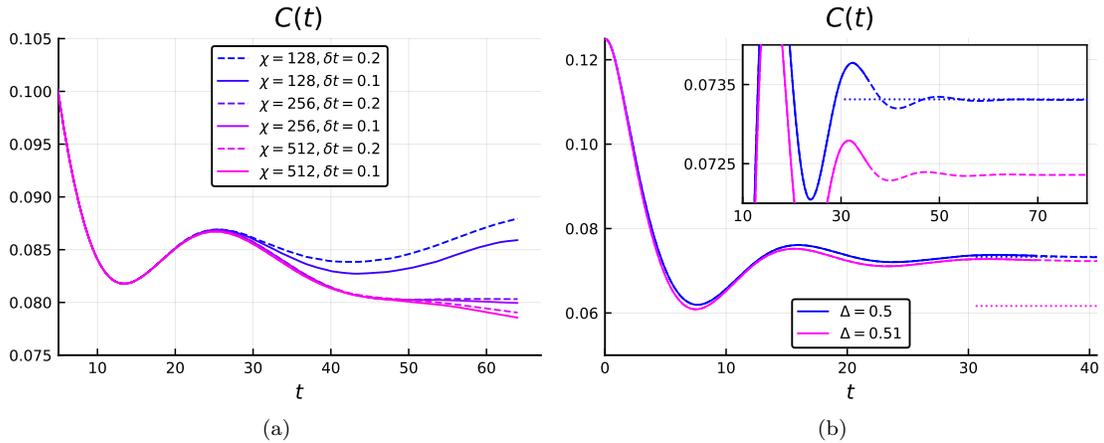

FIG. 1: (a) Convergence of TEBD with increasing bond dimension and decreasing time step in the computation of $C(t)$ for $\lambda = 2/5$. (b) Linear prediction, illustrated using a TEBD computation ($\chi = 512$) of $C(t)$ for $\Delta = 0.5$ and $0.51$. The dotted lines mark the value of the Drude weight for each $\Delta$, which should be the infinite time limit of $C(t)$. The dashed lines show continuation of the time series using a linear prediction technique.

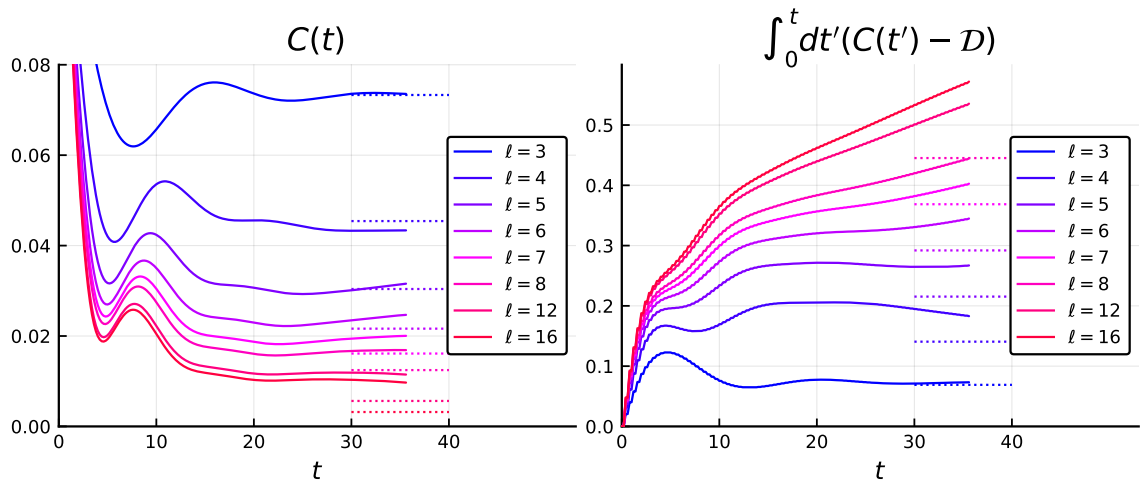

FIG. 2: (a) TEBD computation ($\chi = 512$) of the current-current correlator $C(t)$ of the XXZ chain with $\lambda = \frac{1}{\ell}$, $\Delta = \cos \pi \lambda$. Dashed lines mark the theoretical infinite time limit of $C(t)$ given by the Drude weight $\mathcal{D}$. (b) Integrated correlator $\int_0^t dt' \left( C(t') - \mathcal{D} \right)$ compared to the theoretical infinite time limit given by $\sigma^{\mathrm{d.c.}}$ (dotted lines).

prediction produces a long time limit for $C(t) \approx 0.073307$ that matches the Drude weight $D \approx 0.073313$ with excellent accuracy. At the nearby value $\Delta = 0.51$ (corresponding to an irrational value of $\lambda \approx 30/91$), $C(t)$ is expected to decay to a Drude weight far from $C(t)$ on accessible TEBD time scales. This decay is undetected by the linear prediction, which suggests that it occurs on too large a time scale to significantly effect the TEBD data. As concluded in the main text, large time scales plague the XXZ chain for almost all values of $\Delta$ in the gapless regime - except those corresponding to rational $\lambda$ with small denominators - and thus linear prediction isn't useful for continuing the curves.

To determine whether TEBD has reached the largest time scales significant for a rational $\lambda$, we can compare the TEBD curves with two theoretical predictions - the long time extrapolation of $C(t)$, given by the Drude weight $\mathcal{D}$, and the d.c. conductivity, given by

$$ \sigma^{\mathrm{d.c.}} = \int_0^\infty dt' \left( C(t') - \mathcal{D} \right). $$

Figures 2 and 3 show these comparisons for a variety of rational $\lambda$ corresponding to $\lambda = 1/\ell$ and $\lambda = F_{n-1}/F_{n+1}$, respectively.

## II. THERMODYNAMICS BETHE ANSATZ AND GENERALIZED HYDRODYNAMICS

In this section, we briefly summarize formulas from Thermodynamic Bethe Ansatz (TBA) and generalized hydrodynamics (GHD) relevant to compute the d.c. conductivity in the gapless phase of the XXZ spin chain.

### A. Quasiparticle (string) content

As in the main text, we write the anisotropy as $\Delta = \cos \gamma \equiv \cos \pi \lambda$. The quasiparticle ("strings") spectrum that follows from the Bethe ansatz solution is determined by the continued fraction expansion of $1/\lambda \equiv P_0 = \nu_1 + \frac{1}{\nu_2 + \frac{1}{\nu_3 + \cdots}}$ or equivalently $P_0 = [\nu_1; \nu_2, \nu_3, \ldots]$. The total number of strings is given by $n = \sum \nu_i$. For an irrational value of $P_0$, the number of strings is infinite, while approximating it by a rational number leads to a finite number of strings. In what follows, we follow the standard reference 3. As in the main text, we focus on the specific case of $\lambda = 1/\varphi^2$, where $\varphi = (1 + \sqrt{5})/2$ is the golden mean. (We have checked that the properties below apply to other generic irrational numbers – exceptions include Liouville numbers as discussed in the main text.) The rational approximants are given by $\lambda_n = 1/[2; 1, 1, \ldots, 2] = \frac{F_{n-1}}{F_{n+1}} (n \geq 4)$, where $F_n$ denotes the n$^{\mathrm{th}}$ Fibonacci number. The number of strings for the rational approximant $\lambda_n$ is equal to $n$. The length of the quasiparticles for $\Delta = \cos \pi \lambda_n$ is given by $\ell_j = F_j$ for $i \leq n-2$, $\ell_{n-1} = F_n$, and $\ell_n = F_{n-1}$. The so-called parities of these strings are given by $\kappa_1 = 1$, $\kappa_j = (-1)^{\left\lceil \frac{n_{j-1}}{P_0} \right\rceil}$ for

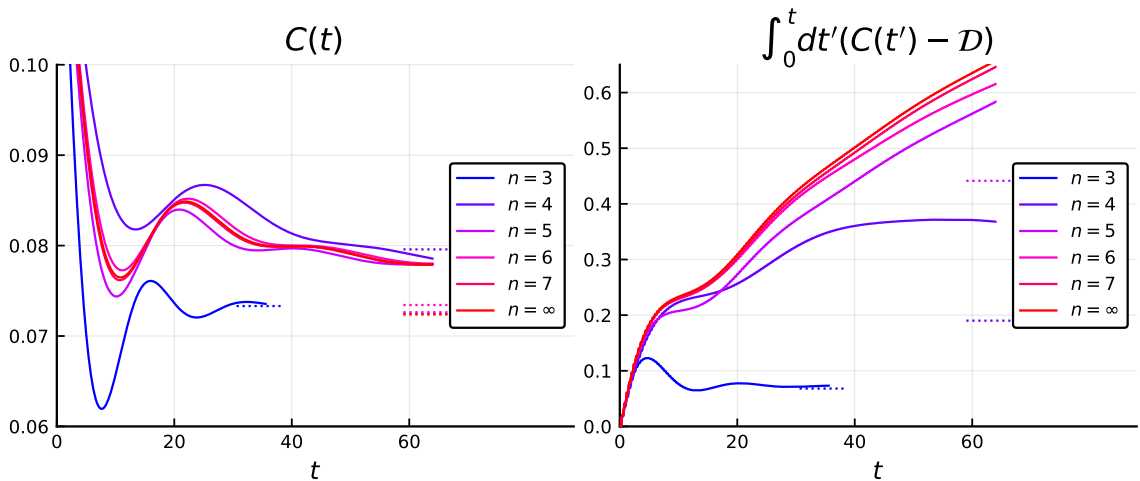

FIG. 3: (a) TEBD computation ($\chi = 512$) of the current-current correlator $C(t)$ of the XXZ chain with $\lambda = \frac{F_{n-1}}{F_{n+1}}$, $\Delta = \cos \pi \lambda$. Dotted lines mark the theoretical infinite time limit of $C(t)$ given by the Drude weight $\mathcal{D}$. (b) Integrated correlator $\int_0^t dt' \left( C(t') - \mathcal{D} \right)$ compared to the theoretical infinite time limit given $\sigma^{\text{d.c.}}$ shown by dotted lines. (For $n > 5$, $\sigma^{\text{d.c.}}$ is above the scale of the plot.)

$j \geq 2$ with the exception of $\kappa_{\nu_1} = -1$. This translates in our specific case of $\lambda_n$ to $\kappa_j = (-1)^{j-1 \bmod 3}$ for $j \leq n-2$, $\kappa_{n-1} = (-1)^{n-1 \bmod 3}$, $\kappa_n = (-1)^{(n-1)-1 \bmod 3}$. Note that the properties of the last two strings are swapped from the general pattern.

## B. Thermodynamic quantities

The equilibrium properties in a thermal state at temperature $T$ are then given by the TBA equations [3]. Let the particle density of string of type $j$ be $\rho_j(\theta)$ and hole density be $\rho_j^h(\theta)$, where $\theta$ is the rapidity. The thermodynamic limit of the Bethe equations read

$$\rho_j^{\text{tot}}(\theta) = \frac{p_j'(\theta)}{2\pi} - \sum_k \sigma_k \int \mathcal{K}_{j,k}(\theta - \alpha) \rho_k(\alpha) d\alpha, \tag{1}$$

where $\sigma_j = \text{sgn}(a_j)$, $p_j(\theta)$ is the momentum (given more explicitly below), $\mathcal{K}_{j,k}$ is the scattering kernel between quasiparticles, and the function $a_j$ is given by

$$a_j(\theta) = \frac{\kappa_j \lambda}{2} \frac{\sin(\pi \lambda \ell_j)}{\cosh(\pi \lambda \theta) - \kappa_j \cos(\pi \lambda \ell_j)}. \tag{2}$$

For an explicit expression of the kernel, we refer the reader to Ref. 3. Minimizing the free energy gives the following TBA equations

$$\ln \eta_j(\theta) = \frac{-2 \sin(\pi \lambda)}{\lambda} \frac{a_j(\theta)}{T} + \sum_k \sigma_k \mathcal{K}_{j,k} \star \ln(1 + \eta_k^{-1}), \tag{3}$$

where $T$ is the temperature, $\star$ denotes the convolution operation, and $\eta_j = \frac{\rho_j^h}{\rho_j}$. The so-called generalized Fermi factor is then given by, $\Theta_j = 1/(1 + \eta_j)$. Combined with the Bethe equations (1), the TBA equations (3) fully specify the quasiparticle densities $\rho_j$ and $\rho_j^h$. In the following, we will focus on the infinite temperature limit.

Due to the interactions among various quasiparticle excitations, conserved charge densities get "dressed":

$$h_j^{\text{dr}}(\theta) = h_j(\theta) - \sum_k \sigma_k \int \mathcal{K}_{jk}(\theta - \alpha) \Theta_k(\alpha) h_k^{\text{dr}}(\alpha) d\alpha, \tag{4}$$

where $h_j(\theta)$ is the bare charge density carried by the $j^{\text{th}}$ string with rapidity $\theta$. Here $h$ could be any conserved charge (magnetization, energy *etc.*). This dressing operation can be dramatic: while the bare magnetization of all strings is

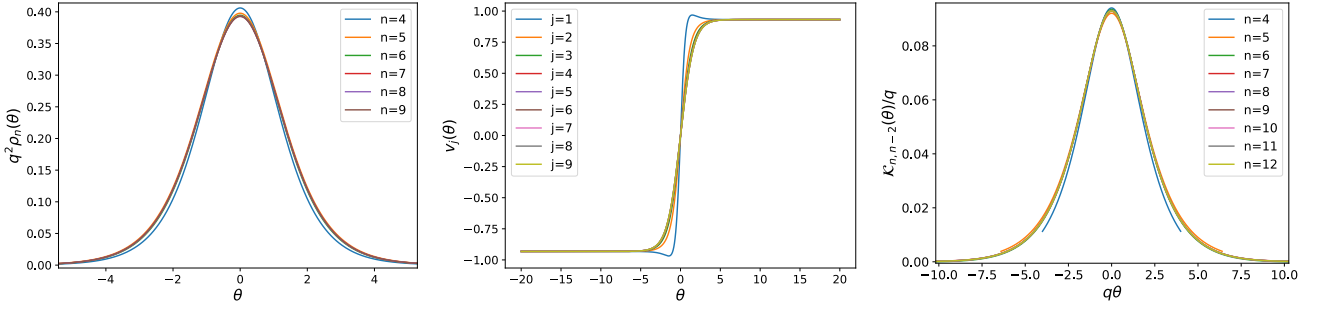

FIG. 4: Scaling of TBA quantities for the Fibonacci series $\lambda_n = \frac{F_{n-1}}{F_{n+1}}$, with $q = F_{n+1}$. Left: the density of the largest string scales as $q^{-2}$. Middle: The velocity saturates for large string numbers. Right: The dominant scattering kernel between the largest neutral string and the two charged strings is sharply peaked, following the scaling (9).

given by their size $m_j = \ell_j$; at infinite temperature, all strings are effectively neutral $m_j^{\mathrm{dr}} = 0$ ($j \leq n-2$) except the largest two which have dressed magnetization $m_j^{\mathrm{dr}} = q/2$ for $j = n-1, n$.

Finally, the effective velocity of the strings in a given equilibrium state is given by [4–6]

$$v_j = \frac{(e'_j)^{\mathrm{dr}}}{(p'_j)^{\mathrm{dr}}},$$ (5)

where $(e'_j)^{\mathrm{dr}}$ and $(p'_j)^{\mathrm{dr}}$ denote the dressed derivatives of energy and momentum, respectively; with the bare values $e'_j = \frac{-2\sin \pi\lambda}{\lambda} a'_j$ and $p'_j = 2\pi a_j$.

## C. Transport coefficients

In terms of the quantities introduced above, the spin Drude weight is given by [7, 8]

$$\mathcal{D} = \sum_k \int d\theta \rho_k(\theta)(1 - \Theta_k)v_k(\theta)^2(m_k^{\mathrm{dr}})^2.$$ (6)

Only the last two (charged) strings contribute to the Drude weight. At infinite temperature, this expression agrees with the more explicit form quoted in the main text. Meanwhile, the d.c conductivity is given by the more complicated expression [9, 10]

$$\sigma^{\mathrm{d.c.}} = \frac{1}{4} \sum_{k,l} \int d\theta_1 d\theta_2 \rho_k(\theta_1)(1 - \Theta_k)\rho_l(\theta_2)(1 - \Theta_l)|v_k(\theta_1) - v_l(\theta_2)| \left(\mathcal{K}_{kl}^{\mathrm{dr}}(\theta_1, \theta_2)\right)^2 \left(\frac{m_k^{\mathrm{dr}}\sigma_k\Theta_k}{\rho_k(\theta_1)} - \frac{m_l^{\mathrm{dr}}\sigma_l\Theta_l}{\rho_l(\theta_2)}\right)^2,$$ (7)

where $K_{ij}^{\mathrm{dr}}$ is the dressed scattering kernel, given by

$$\mathcal{K}_{ij}^{\mathrm{dr}}(\theta, \alpha) = \mathcal{K}_{ij}(\theta, \alpha) - \sum_k \sigma_k \int d\beta \mathcal{K}_{ik}(\theta, \beta)\Theta_k \mathcal{K}_{kj}^{\mathrm{dr}}(\beta, \alpha).$$ (8)

## D. Scaling of the scattering kernels and conductivity

Focusing on the Fibonacci series $\lambda_n = \frac{F_{n-1}}{F_{n+1}}$, all the above quantities exhibit a simple scaling form in terms of $q = F_{n+1}$ as $n \to \infty$ (Fig. 4). First of all, the velocities quickly saturate to a $q$-independent form for string labels $j \gg 1$. On the other hand, it is easy to show that the density of the charged quasiparticles $j = n-1$ and $j = n$ scales as $\rho_n \sim q^{-2}$. This scaling form is necessary for the Drude weight (6) to be finite, since $m^{\mathrm{dr}} \sim q$ for the large, charged strings. From the expression of the scattering kernel, it is straightforward to see that the dominant scattering processes

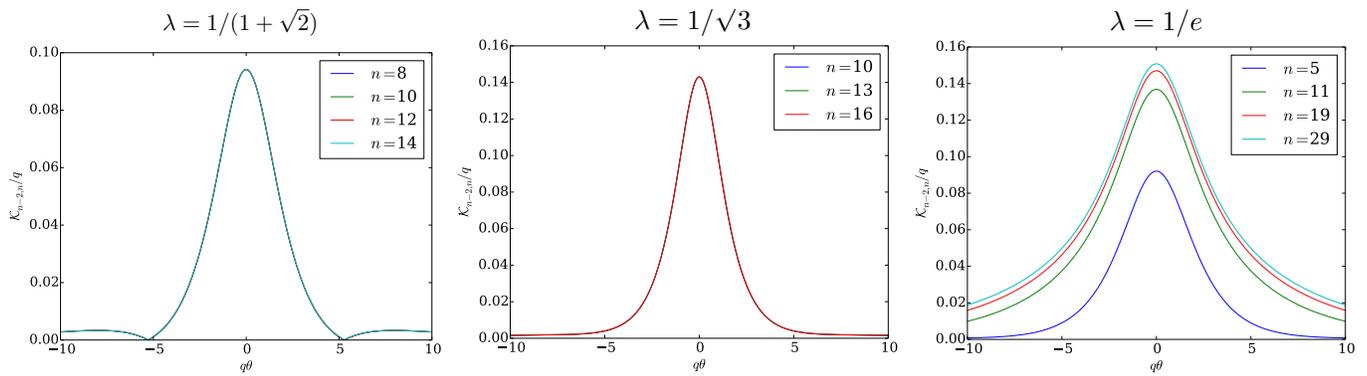

FIG. 5: Scaling of the absolute value of the scattering kernel between the largest neutral quasiparticle $j = n - 2$ and one of the two charged quasiparticles $j = n$ for the rational approximations (with denominator $q$) of various irrational numbers. For irrationals with periodic continued fraction expansion such as $1/(1+\sqrt{2})$ or $1/\sqrt{3}$, the kernel collapses perfectly as in the golden ratio case. For irrationals such as $\lambda = 1/e$ whose continued fraction expansion is aperiodic, there is no reason to expect the kernel to collapse onto a single form. However, the kernel still becomes sharply peaked as the number of quasiparticles $n \to \infty$: the width of the peak very clearly scales as $q^{-1}$, while the height of the peak diverges as $q$, possibly plus some small aperiodic fluctuations.

for the charged strings are with the largest neutral strings. The (bare) scattering kernel $\mathcal{K}_{n-2,n}(\theta, \alpha) = \mathcal{K}_{n-2,n}(\theta - \alpha)$ for such strings becomes increasingly peaked near $\theta \to 0$ as $n \to \infty$

$$\mathcal{K}_{n-2,n}(\theta) = q f(q\theta), \tag{9}$$

where the function $f$ depends on the irrational number that is being approximated. Meanwhile, the scattering kernel of the charged strings with light (small) neutral strings is strongly suppressed, with a peak that scales as $q^{-2}$ and a $q$-independent width.

The scaling (9) can be seen as follows: the scattering kernel is a sum of $a(\theta)$ functions with $\ell_j$ being replaced by sums and differences of particle lengths [3]. For the dominant scattering kernel $\mathcal{K}_{n-2,n}(\theta)$, by inspection, we find that one of the two dominant terms in this sum corresponds to an effective length $F_n$, with a contribution which becomes sharply peaked as claimed:

$$\frac{\sin(\pi\lambda F_n)}{\cosh(\pi\lambda\theta) - \kappa_n \cos(\pi\lambda F_n)} \sim \frac{\pi\sqrt{5}\varphi^{-n-1}}{\cosh(\pi\lambda\theta) - 1 + 5\pi^2\varphi^{-2n-2}/2} \sim \frac{2\varphi}{\pi\sqrt{5}} \times \frac{\varphi^n}{1 + (\theta\varphi^{n-1}/\sqrt{5})^2}, \tag{10}$$

which indeed has a peak of order $q \sim \varphi^n$, and width $1/q$. We have used the fact that $\lambda F_n$ is almost an integer, so that $\sin(\pi\lambda F_n) \sim \pi\sqrt{5}\varphi^{-n-1}$ is exponentially small in $n$. This requires taking the rapidity $\theta \to 0$ with $\theta\varphi^n$ fixed to get a non-zero contribution. The other dominant term in $\mathcal{K}_{n-2,n}(\theta)$ behaves similarly, so the bare kernel does indeed scale as (9).

Using the fact that the Fermi factor is independent of the rapidity at infinite temperature, one can show from (8) that the dressed kernel obeys the same scaling form (9) as $n \to \infty$. Since the (leading) bare kernels scale as (9), it is easy to see that this scaling property is preserved by the convolution: $\int d\beta \mathcal{K}(\theta - \beta) qg[q(\beta - \beta)] = q \int d(q\beta) f[q(\theta - \beta)] g[q(\theta - \alpha)] = qh[q(\theta - \alpha)]$ with $h = f \star g$. Solving the dressing equation (8) by iteration, this implies that this scaling property is preserved by the dressing.

These scaling forms are enough to argue that the charged particles undergo Lévy flights as explained in the main text. From eq. (9), the scaling of the d.c. conductivity (7) is also clear. Since only the last two strings are charged, we can set $k = n$ (or $k = n - 1$) leaving a single sum over quasiparticles. Consider the contribution from the scattering with the largest neutral string $l = n - 2$: the density factors contribute $q^{-4}$, and $(m_n^{dr}/\rho_n)^2 \sim q^6$ since the Fermi factors remain $\mathcal{O}(1)$. The square of the dressed kernel scales as $(\mathcal{K}^{dr})^2 \sim q^2$, but it forces $|\theta_1 - \theta_2| < q^{-1}$ since it is sharply peaked: this gives a $q^{-1}$ factor from the integral $\int d(\theta_1 - \theta_2)$, as well as an additional $q^{-1}$ from $|v_k(\theta_1) - v_l(\theta_2)|$ since the velocities are forced to be almost the same up to $q^{-1}$. Gathering these pieces, this contribution scales as $q^{-4}q^6q^2q^{-4} \sim q^2$. If we consider instead the scattering with a light neutral string such as $l = 1$, one of the density factors is now $\rho_k \sim \mathcal{O}(1)$, but the square of the dressed kernel scales as $q^{-4}$, with a width that does not scale with $q$ so there is no factor from the integration interval, and $|v_k(\theta_1) - v_l(\theta_2)| \sim \mathcal{O}(1)$. This gives a contribution of order $q^{-2}q^6q^{-4} \sim 1$, so it is much smaller than the contribution from the scattering due to heavy (large) neutral strings. We thus conclude that

$$\sigma_{\lambda_n}^{d.c.} \sim q^2 \sim \varphi^{2n}. \tag{11}$$

These scaling properties remain valid for generic irrational numbers as illustrated in Fig 5.

---

[1] C. Karrasch, D. M. Kennes, and J. E. Moore, Phys. Rev. B **90**, 155104 (2014), URL https://link.aps.org/doi/10.1103/PhysRevB.90.155104.

[2] C. Karrasch, D. M. Kennes, and F. Heidrich-Meisner, Phys. Rev. B **91**, 115130 (2015), URL https://link.aps.org/doi/10.1103/PhysRevB.91.115130.

[3] M. Takahashi, *Thermodynamics of One-Dimensional Solvable Models* (Cambridge University Press, 1999), ISBN 9780521551434, URL https://books.google.com/books?id=kX1FAwEACAAJ.

[4] L. Bonnes, F. H. L. Essler, and A. M. Läuchli, Phys. Rev. Lett. **113**, 187203 (2014), URL https://link.aps.org/doi/10.1103/PhysRevLett.113.187203.

[5] B. Bertini, M. Collura, J. De Nardis, and M. Fagotti, Phys. Rev. Lett. **117**, 207201 (2016), URL http://link.aps.org/doi/10.1103/PhysRevLett.117.207201.

[6] O. A. Castro-Alvaredo, B. Doyon, and T. Yoshimura, Phys. Rev. X **6**, 041065 (2016), URL http://link.aps.org/doi/10.1103/PhysRevX.6.041065.

[7] E. Ilievski and J. De Nardis, Phys. Rev. B **96**, 081118 (2017), URL https://link.aps.org/doi/10.1103/PhysRevB.96.081118.

[8] B. Doyon and H. Spohn, SciPost Phys. **3**, 039 (2017), URL https://scipost.org/10.21468/SciPostPhys.3.6.039.

[9] J. De Nardis, D. Bernard, and B. Doyon, Phys. Rev. Lett. **121**, 160603 (2018), URL https://link.aps.org/doi/10.1103/PhysRevLett.121.160603.

[10] J. D. Nardis, D. Bernard, and B. Doyon, SciPost Phys. **6**, 49 (2019), URL https://scipost.org/10.21468/SciPostPhys.6.4.049.



\end{document}